\documentclass[manuscript]{acmart}

\usepackage{algorithmic,algorithm}
\usepackage[utf8]{inputenc} 
\usepackage{url}            
\usepackage{amsfonts}       
\usepackage{nicefrac}       
\usepackage{microtype}      
\usepackage{algorithmic,algorithm}

\usepackage{xspace}
\usepackage{subfigure}

\newtheorem{theorem}{Theorem}

\newtheorem{proposition}{Proposition}

\newcommand{\bed}{${\lambda}\texttt{BED}$\xspace}
\newcommand{\red}{${\lambda}\texttt{RED}$\xspace}

\newcommand{\pbed}{$\texttt{BED}$\xspace}
\newcommand{\pred}{$\texttt{RED}$\xspace}
\newcommand{\prob}{\textsc{Paegs}\xspace}
\newcommand{\sprob}{\textsc{Paegs-b}\xspace}

\AtBeginDocument{%
  \providecommand\BibTeX{{%
    \normalfont B\kern-0.5em{\scshape i\kern-0.25em b}\kern-0.8em\TeX}}}

\acmYear{2019}

\acmBooktitle{}
\acmPrice{15.00}
\acmISBN{}

\begin{document}

\title{Learning-Assisted Competitive Algorithms for Peak-Aware Energy Scheduling}

\author{Russell~Lee}
\email{rclee@cs.umass.edu}
\affiliation{%
  \institution{University of Massachusetts Amherst}
}

\author{Mohammad H. Hajiesmaili}
\email{hajiesmaili@cs.umass.edu}
\affiliation{%
	\institution{University of Massachusetts Amherst}
}

\author{Jian~Li}
\email{lij@binghamton.edu}
\affiliation{%
	\institution{Binghamton University, the State University of New York}
}

\renewcommand{\shortauthors}{Lee, Hajiesmaili, and Li}

\begin{abstract}
In this paper, we study the peak-aware energy scheduling problem using the competitive framework with machine learning prediction. With the uncertainty of energy demand as the fundamental challenge, the goal is to schedule the energy output of local generation units such that the electricity bill is minimized. While this problem has been tackled using classic competitive design with worst-case guarantee, the goal of this paper is to develop learning-assisted competitive algorithms to improve the performance in a provable manner. We develop two deterministic and randomized algorithms that are provably robust against the poor performance of learning prediction, however, achieve the optimal performance as the error of prediction goes to zero. Extensive experiments using real data traces verify our theoretical observations and show 15.13\% improved performance against pure online algorithms.
\end{abstract}



\maketitle

\section{Introduction}
\label{sec:intro}
The electricity bill is a significant operating cost of large energy customers such as data centers, business headquarters, and university campuses. For example, the largest expenditure to operate a data center is the energy consumption, e.g., with more than 30\% of total operating costs of Google and Microsoft's data centers~\cite{qureshi2009cutting}. Consequently, managing the energy consumption and cost of large energy costumers such as data centers has become critically important and there have been substantial research in this direction on incorporating local renewable sources~\cite{liu2012renewable}, energy-aware server provisioning~\cite{lin2013dynamic}, geographical load balancing~\cite{le2018optimal,liu2014greening}, and on-site energy storage systems~\cite{yang2019online,hajiesmaili2017crowd,ahmad2019learning}.

The electricity bill for large energy customers is usually a hybrid volume and peak-based charging model. Specifically, considering a time slotted model of length $T$ slots as the billing cycle, and the energy demand $d(t)$ in slot $t$, the electricity bill is the aggregation of the following two terms: (1) the \textit{volume pricing}, which is the aggregate energy usage over the cycle, i.e., $\sum_{t} p(t)d(t)$, with $p(t)$ as the real-time unit price at $t$, and (2) the \textit{peak pricing}, which is the peak demand drawn over the cycle, i.e., $ \max_{t \in [T]} d(t)$, multiplied by $p_{m}$ as the peak price. 
The contribution of peak pricing in the electricity bill is usually substantial. The peak price is often more than 100 times higher than the maximum spot price, e.g., 118 times for PG\&E, and 227 times for Duke Energy Kentucky. Hence, the contribution of peak in the energy bill for large energy costumers can be considerable, e.g., from 20\% to 80\% for Google data centers~\cite{xu2013reducing}. 

A promising approach to reduce the contribution of peak charging in the final electricity bill is to install on-site generation units, and cover a portion of the demand to shave the peak~\cite{zhang2018peak}. Notable examples are Google data center in Council Bluffs, Iowa with 38 local generators, and Microsoft's plan to add 72 new generators at its Quincy, Washington, data center campus. The global market of data center generator is expected to reach the revenue of around \$5 billion in 2023. With the on-site generator, one can schedule its generation such that part of the total energy demand is satisfied by the local generator, hence, the peak net demand from the grid is reduced over the billing cycle. 

Peak-aware energy generation scheduling of local generator, however, is a challenging problem due to the uncertainty of the demand of energy customers, especially data centers. For example, for data centers the energy demand is highly unpredictable because user demand for internet services is extremely variable. For instance, a data center serving videos to users can experience an unexpected flash crowd of users for a popular video release. Further, sophisticated optimization algorithms are being used in Google data centers to improve the energy efficiency of data center's internal operations~\cite{DeepMind}, which can further increase the unpredictable variability of energy demand. Also, in geographical load balancing schemes~\cite{le2018optimal,liu2014greening}, a load balancer could move user demand into another data center, resulting in unexpected changes in the energy patterns.  Last, the integration of renewables into data centers is another uncertainty, since production level of renewable is uncertain and intermittent~\cite{goiri2013parasol}.  

The peak-aware energy generation scheduling problem (\prob, henceforth) has been tackled using the competitive online framework~\cite{zhang2018peak}. More specifically, two deterministic and randomized algorithms have been proposed that can achieve the best \textit{competitive ratio} as the well-established performance metric for online algorithms~\cite{Borodin98}. Competitive ratio is defined as the ratio between the cost of an online algorithm and that of offline optimal, worst-case over all feasible instances to the problem. The competitive online framework, however, in its basic form aims to be provably efficient against worst-case realizations of input instances. Toward this, it assumes that no stochastic, exact, or noisy measurement of future data is available and tries to make best decisions without the future knowledge. Trying to be efficient against worst-case makes competitive algorithms to be conservative and pessimistic in common realization of inputs since worst-case scenarios are usually instances that rarely happen in reality. On top of that, in most online problems including \prob, it is reasonable to have at least a noisy prediction of future data.

This goal of this paper is to design learning-assisted competitive algorithms for \prob. Our approach is inspired by the recent effort on integrating machine learning (ML) predictions to improve the performance of online algorithms in theory and practice~\cite{purohit2018improving,lykouris2018competitive,kodialam2019optimal,gollapudi2019online}. The key motivation is two-fold: (1) to keep the core competency of online algorithms, i.e., performance guarantee against worst-case; and (2) to achieve a provably improved performance if the accuracy of ML-predictor is satisfactory. The two key motivations could be theoretically analyzed for the \textit{learning-assisted online algorithms}~\cite{purohit2018improving,lykouris2018competitive} by introducing the notions of (1) \textit{robustness} that characterizes the first motivation; and (2) \textit{consistency} that characterizes the second one.
Specifically, suppose that $\mathcal{A}$ is a learning-assisted online algorithm  that leverages an ML-predictor in decision making. The algorithm $\mathcal{A}$ is $(\alpha,\gamma)$-competitive where $\alpha$ and $\gamma$ represent the robustness and consistency of $\mathcal{A}$, respectively. That is, the competitive ratio of $\mathcal{A}$ is always less than $\alpha$ regardless of the error in ML-predictor. Also, $\mathcal{A}$ is \emph{$\gamma$-consistent} if with perfect ML prediction it achieves the competitive ratio of $\gamma.$  Robustness measures how well the algorithm does in the worst-case of poor prediction, and consistency measures how well the algorithm does under perfect prediction.  

With this new analytical framework, one is able to achieve ``the best of both world'' paradigm from the perspective of learning-assisted competitive algorithms. While it might slightly degrade the robustness against worst-case, or ideally maintains the worst-case guarantee, it resolves the fundamental drawback of competitive analysis of pessimistic decision making with incorporating ML predictions. More importantly, different from the classic prediction-based competitive designs~\cite{chen2015online,chen2016using,boyar2016online,hajiesmaili2016online}, the framework used in this paper leverages a hyperparameter that determines how much the algorithmic decisions trust on the predictors, enabling the full spectrum coverage from pure worst-case to fully prediction-based decision making.

\subsubsection{Summary of contributions:} Inspired by the above direction of learning-assisted competitive design, we develop two deterministic and randomized algorithms for \prob that take into account the noisy prediction from a machine learning model in decision making, and improve the performance of existing pure online algorithms in a provable manner. 

First, we propose \bed, a deterministic algorithm parameterized by a hyperparameter $\lambda \in (0,1)$, that achieves a competitive ratio of $1 + (1-\beta)/\lambda,$ where $\beta\in(0,1]$ is a problem specific parameter.  In particular, \bed is $(1 + (1-\beta)/\lambda)$-robust and $(1 + {\lambda})$-consistent. The trust on ML prediction is interpreted as follows. Greater trust in ML prediction is achieved by setting $\lambda$ close to zero, which means that the algorithm can achieve the optimal performance. On the other hand, less trust on ML advice is achieved by setting $\lambda$ close to one, and the robustness results guarantees the same performance with \pbed~\cite{zhang2018peak} as the pure online algorithm with the optimal competitive ratio. 

Second, we propose \red, a randomized algorithm with hyperparameter $\lambda$ that
is $\left(\frac{1}{e-1+\beta} \left[e + \frac{(1-\lambda)(1-\beta)(e-1+\beta)}{\beta}\right]\right)$-robust and $\left( \frac{1}{e-1+\beta}\left[ e+(\lambda -1)(1-\beta) + \frac{\lambda(1-\lambda)(1-\beta)(e-1)}{\beta }\right] \right)$-consistent. With $\lambda=1$, \red recovers the competitive ratio of the best online algorithm, and with $\lambda=0$ it behaves optimally. 
Design and competitive analysis of \red is the significance of theoretical contributions of this paper. Specifically, it is worth noting that the probability distribution functions of \red are carefully designed to achieve a solid robustness and consistency guarantees. This distribution function is customized based on Yao's principle~\cite{yao1977probabilistic} and provides high-level insights for developing online randomized algorithms, hence, it provides the robustness and consistency results in a more systematic manner as compared to the randomized algorithm design for online problems in~\cite{purohit2018improving,lykouris2018competitive,kodialam2019optimal}. Further, we show that the straightforward extension of existing randomized algorithm fails to guarantee solid performance. 


Last, we empirically evaluate the performance of the algorithms using real-world data traces. We use the energy demand traces from Akamai data centers that serve about a quarter of the Web traffic worldwide~\cite{nygren2010akamai}. We use energy price values from New York energy market (NYISO). The results show the improved performance of the proposed learning-assisted algorithm as compared to the pure online algorithm. As a representative experimental result, \bed achieves 15.13\% cost reduction as compared to \pbed. We also investigate the impact of several parameters and provide several insights that reveal the benefits of learning-assisted algorithms in practice. 

\section{Problem Formulation and Existing Algorithms}
\label{sec:prob}

\subsubsection{Problem statement:} The goal of the peak-aware energy generation scheduling problem (\prob) is to minimize the energy cost over the billing cycle while satisfying the electricity demand.  We consider one billing cycle $\mathcal{T}=\{1,\cdots, T\}$ with $T<\infty$ discrete time slots of uniform length. The billing cycle is usually one month and the length of each slot is one hour.  Let $d(t) \geq 0$ be the electricity demand in slot $t$. The values of demand are known for current and previous slots, but, not known for future slots. The demand could be covered from two sources, \emph{local generators} and \emph{the external grid}.   The local generators can satisfy at most $C\geq 1$ KW of energy demand in each slot, with the unit cost $p_g.$ In reality, some traditional generators have maximum ramp-up and ramp-down constraints that limits the change of output in two slots. In the solution section of this paper, we focus on ``fast-responding" generators that are fast enough to ramp up and down without any limit. In experiments, however, we investigate the impact of the ramp constraints.

Following the dynamics of the energy market, the grid provides electricity with a spot price $p(t)$ at time $t,$ where we assume $p(t)\geq p^{\min}>0.$ In reality, the unit cost of local generators is usually higher than that of external grid, i.e., $p_g\geq p(t)$. Otherwise, it is always optimal to use local generators as much as possible for both online and offline algorithms.  However, the expensive local generator can cut off the peak demand (peak charge) from the external grid. In addition, $p_m$ is the peak charge price that is known and fixed over the billing cycle. Note that $p_m$ is usually more than $100$ times larger than $p(t)$.  
For ease of exposition, denote $\beta\triangleq p^{\min}/p_g < 1$ as the ratio between the minimum grid price and the unit cost of local generation. We characterize the performance of our algorithms as a function of $\beta$.

Let $v(t)$ and $u(t)$ be the optimization variables that determine the amount of electricity procured from the external grid and local generator, respectively.  For the grid, its cost consists of volume charge and peak charge. The volume charge is the sum of volume cost over the time horizon, i.e., $\sum_t p(t) v(t).$  The peak charge is based on the maximum single-slot power and peak price $p_m,$ i.e., $p_m\max_t v(t)$ \cite{xu2013reducing,zhang2018peak}.   The cost of using local generators, is $\sum_t p_g u(t).$    Therefore, with $\boldsymbol{u} = [u(t)]_{t\in\mathcal{T}}$ and $\boldsymbol{v} = [v(t)]_{t\in\mathcal{T}}$, the total operating cost over the billing cycle is

\begin{align*}
\textsc{Cost}(\boldsymbol u, \boldsymbol v)=\sum_{t\in\mathcal{T}}p(t) v(t)+p_m\max_t v(t)+\sum_{t\in\mathcal{T}}p_g u(t).
\end{align*}
The \prob problem is defined as follows,
\begin{align*}
\prob: &\quad\min_{\boldsymbol u, \boldsymbol v}\quad \textsc{Cost}(\boldsymbol u, \boldsymbol v)
\\
&\text{s.t.,}\quad u(t)+v(t)\geq d(t),  \quad  
u(t)\leq C, \quad t\in\mathcal{T},
\end{align*}                        
where the first constraint ensures that the electricity demand is satisfied, and the second constraint is due to the generator capacity limitation. 
In the offline setting, where $d(t)$ is fully known in advance, \prob can be solved using any general algorithm for linear programming.  However, in practice, the demand $d(t)$ is hard to predict, hence an online algorithm that does not rely on demand prediction is preferred. In the following, we briefly review existing algorithms for \prob using competitive framework. In the next section, we develop two algorithms that  integrate machine learning predictions to design online algorithms that are both robust and consistent.

\subsubsection{Summary of prior work:} In prior work~\cite{zhang2018peak} online algorithms are developed to solve \prob using competitive framework~\cite{Borodin98}. The key is to construct a basic version of \prob first, named as \sprob, where the net demand only takes values $0$ or $1$, and then extend it to the general \prob. Note that the procedure for generalization of the algorithms from the basic version to the general case applies to the proposed algorithms in this paper as well. Hence, hereafter we focus on solving \sprob. The \sprob is defined as follows, 
\begin{align*}
&\sprob: \min\limits_{\boldsymbol u, \boldsymbol v} \textsc{Cost}(\boldsymbol u, \boldsymbol v)  \\ 
& \text{s.t.,}\quad u(t)+v(t)\geq d(t), u(t), v(t)\in\{0, 1\}, t\in\mathcal{T}.
\end{align*} 
In the following, we first recall the algorithms in~\cite{zhang2018peak}. 
The key in solving \sprob lies in balancing between the cost of using expensive local generators and the peak charge of using the external grid.  

\noindent\textit{An Optimal Offline Algorithm \emph{\cite{zhang2018peak}}.} The key in algorithm design for \sprob is to define $\sigma$ as the critical  peak-demand threshold as 
\begin{equation}
\label{eq:sigma}
\sigma=\frac{1}{p_m}\Big[\sum_{t\in\mathcal{T}}(p_g-p(t))d(t)\Big].
\end{equation}
The parameter $\sigma$ plays a critical role in algorithm design. For optimal offline algorithm, we have $v^*(t)= d(t)$, $\forall t\in\mathcal{T}$, when $\sigma>1;$ and $v^*(t)= 0$, $\forall t\in\mathcal{T}$, otherwise. The optimal local generator output is then $u^*(t)=d(t)-v^*(t).$

\noindent\textit{A deterministic online algorithm \emph{(\pbed)}.} The value of $\sigma$ could be calculated easily in offline manner, however, with unknown price and demand values, this values cannot be computed in online setting. The high-level idea of \pbed is to make decisions based on the calculated value of $\sigma$ over the current and past slots. Specifically, \pbed keeps using the local generator initially and switches to the grid at the first time $\tau$ such that 
$\sum_{t=1}^\tau (p_g - p(t))d(t) \geq  p_m.$
The competitive ratio of \pbed is $2-\beta$.  
Similar to the ski-rental problem, the break-even point is the best balance between being aggressive (paying the one-time premium peak cost) and conservative (on using local generator). 

\noindent\textit{A randomized online algorithm \emph{(\pred)}.}  \pred randomly selects $s$ to start purchasing grid electricity when $\sum_{\tau}(p_g - p(\tau)) \geq s \cdot p_m$ according to the following distribution
\begin{align}\label{eq:dis-zhang}
f^*(s)=  \begin{cases} 
\frac{e^s}{e-1+\beta}, & \text{when }s \in [0,1]; \\
\frac{\beta}{e-1+\beta} \delta(0), &\text{when }s = \infty; \\
0, & \text{otherwise}.
\end{cases}
\end{align}
The competitive ratio of \pred is $e/(e-1+\beta).$ Here, $f^*(s)$ is the same distribution as used in solving the classic Bahncard problem \cite{fleischer2001bahncard}, however, the price $p(t)$ is time varying in our problem. 


\section{Learning-Assisted Online Algorithms}
\label{sec:alg}

In this section, we develop a deterministic algorithm, called \bed, and a randomized algorithm, called \red. Both algorithms enhance the practical performance of \pbed and \pred by integrating predictions from machine learning in decision making, while keeping their competitiveness in worst-case.

\subsection{\bed: A Deterministic Algorithm}
%
%

We develop a new deterministic algorithm by adding a hyperparameter $\lambda \in (0,1)$ that facilitates incorporating ML prediction actions and analyze its consistency and robustness as introduced in Introduction. 

First, we introduce the additional input as the result of ML prediction. Assume that there is a learning model that predicts the future values of external grid prices, $\hat{p}(t)$, and energy demand, $\hat{d}(t)$. We do not assume any modeling from machine learning and treat is as a black-box that provides input to our algorithms. Given these two values, let $\hat{\sigma}$ be the predicted critical peak-demand threshold
\begin{align}
\label{eq:sigma_hat}
\hat{\sigma} \triangleq \frac{1}{p_m}\Big[\sum_{t \in \mathcal{T}}(p_g - \hat{p}(t))\hat{d}(t)\Big].
\end{align}
Note that it is even possible that the ML-predictor directly predicts the value of $\hat{\sigma}$ based on historical break-even points in previous cycles. In this way, there is no need to predict individual values of $\hat{p}(t)$ and $\hat{d}(t)$ for the entire cycle. 
Then, the algorithm makes decisions based on the value of $\hat{\sigma}$ and hyperparameter $\lambda$ as summarized in Algorithm~\ref{alg:bed}. 

\begin{algorithm}
	\caption{\bed}
	\begin{algorithmic}
		\label{alg:bed}
		\STATE \textbf{if} $\hat{\sigma} > 1$ \textbf{then} 
		$ s \leftarrow \lambda$
		\textbf{else}
		$s \leftarrow  \frac{1}{\lambda} $
		\textbf{end if}
		\STATE Use local generator first and switch to the grid electricity starting at the first time $\tau$ where
		
		$$\sum\nolimits_{t=1}^\tau (p_g - p(t))d(t) \geq  s \cdot p_m.$$
	\end{algorithmic}
\end{algorithm}
Now, we analyze the robustness and consistency of the \bed algorithm.  Given the general structure of Algorithm~\ref{alg:bed}, we can parameterize any online algorithm by parameter $s$. Let $\mathcal{A}_s$ be an online algorithm with a specific parameter $s$, e.g., \bed is in this category with the value of $s$ as in the first line of Algorithm~\ref{alg:bed}.  Let $h(\mathcal{A}_s,\sigma)$ be the ratio between the cost of  algorithm $\mathcal{A}_s$ and that of an optimal offline algorithm. First, the following proposition characterizes the closed-form value of $h(\mathcal{A}_s,\sigma)$ as a function of $\sigma$ and $s$, and facilitates the analysis of the proposed algorithm in this section.
\begin{proposition}\label{prop:deterministic}
	\emph{\cite{zhang2018peak}} For any online algorithm $\mathcal{A}_s$, we have,
	when $\sigma \leq 1$,
	$$h(\mathcal{A}_s,\sigma) =  \begin{cases} 
	1, & \text{if }s>\sigma; \\
	1 + \frac{1-\sigma + s}{\sigma}(1-\beta), & \text{otherwise}.
	\end{cases}
	$$
	when $\sigma >1$,
	$$h(\mathcal{A}_s,\sigma) =  \begin{cases} 
	1 + \frac{(\sigma-1)(1-\beta)}{(\sigma-1)\beta +1},& \text{if }s> \sigma; \\
	1 + \frac{s(1-\beta)}{(\sigma-1)\beta +1}, & \text{otherwise.}
	\end{cases}
	$$
\end{proposition}

\begin{theorem}\label{thm:deterministic} 
	The \emph{\bed} algorithm achieves the competitive ratio of $1 + (1-\beta)/\lambda,$ where $\lambda\in(0,1)$.  In particular, \emph{\bed} is $(1 + (1-\beta)/\lambda)$-robust and $(1 + {\lambda})$-consistent.
\end{theorem}
\textbf{Remarks.} (1) Setting $\lambda = 1$ in robustness result recovers the competitive ratio of \pbed as the optimal online algorithm. This implies that with bad prediction it suffices to set the value of $\lambda$ to one to be robust against worst-case.  (2) Setting $\lambda = 0$ in consistency results in a competitive ratio of $1$. This captures the case of accurate prediction and implies that with perfect prediction, \bed achieves optimal performance. Hence, by tuning the value of $\lambda$, one can change the importance of prediction from machine learning in decision making. (3) It is worth noting that different from the classic competitive design, in this approach, the competitive ratio is characterized as a function of the hyperparameter $\lambda$. By varying the value of $\lambda$, one can achieve different values for the competitive ratio that for some cases might be even worse than the classic online algorithms, e.g., having $\beta = 0.75$ and setting $\lambda = 0.5$, \bed guarantees the robustness of 1.5 and consistency of 1.5, whereas \pbed guarantees better competitive ratio of 1.25. This shows that  relying on ML-predictors in decision making comes at the expense of lower worst-case performance guarantee as the fundamental trade-off between robustness in worst-case and improving practical performance by incorporating prediction. Last, it signifies that tuning $\lambda$ is the key for improving the performance, e.g., in above example, setting $\lambda=0$ yield better consistency than \pbed, and setting $\lambda=1$ yields the same performance as \pbed.

\textit{Sketch of the proof of theorem~\ref{thm:deterministic}:} We provide the sketch of the proof, and a detailed derivation of the proof is given in the appendix.
We first consider the robustness.    The worst-case cost ratio for a general deterministic algorithm $\mathcal{A}_s$ is achieved with $\sigma=s,$ i.e., the online algorithm pays for the peak charge premium but has no demand to serve anymore.  From Proposition~\ref{prop:deterministic}, the competitive ratio of $\mathcal{A}_s$  is 

\begin{align*}\label{eq:worst-ratio}
\text{CR}(\mathcal{A}_s) \!=\!  \max_{\sigma}h(\mathcal{A}_s,\sigma) \!= \! \begin{cases} 
1 + \frac{1}{s}(1-\beta), \!&\! \text{if }s\leq 1, \\
1 + \frac{s(1-\beta)}{(s-1)\beta + 1}, \!&\! \text{o.w}.
\end{cases}
\end{align*}
We compute the competitive ratio of  \bed under two cases

(i) $\hat{\sigma} >1$: According to \bed, $s = \lambda < 1$. 
From above equation, we have $\text{CR}(\mathcal{A}_{\lambda}) = 1 + (1-\beta)/\lambda.$

(ii) $\hat{\sigma} \leq 1$: According to \bed, $s = {1/\lambda} > 1$. 
From above equation, we have $\text{CR}(\mathcal{A}_{1/\lambda}) = 1 + \frac{(1-\beta)/\lambda}{(1/\lambda-1)\beta + 1}.$

Since $1/\lambda-1 > 0, \beta \geq 0$, we have ${(1/\lambda-1)\beta + 1 \geq 1}$.  We take the overall competitive ratio of \bed to be $\max \{\text{CR}(\mathcal{A}_{1/\lambda}),\text{CR}(\mathcal{A}_{\lambda})\}= 1 + (1-\beta)/\lambda.$ This implies that \bed is $(1 + (1-\beta)/\lambda)$-robust. 
Next, we consider the consistency.  For consistency, we compute the competitive ratio assuming perfect predictions. There are two cases

(i)  $\hat{\sigma} = \sigma >1$, i.e., $s=\lambda.$ With hyperparameter $\lambda$, the algorithm uses the local generator for the first $T^{\lambda}$ time slots before switching to the grid.  Then the cost of \bed  denoted by \text{ALG}, is   ${\text{ALG} = \sum_{t=1}^{T^{\lambda}}p_gd(t) + \sum_{t=T^{\lambda}+1}^{T}p(t)d(t) + p_m}$.  Since $\sigma > 1$, the offline optimal always uses the grid with cost  $\text{OPT} = \sum_{t=1}^{T}p(t)d(t) + p_m$.   Then, we have 
\begin{align*}
\text{ALG} &\stackrel{(a)}{\leq}\lambda \cdot p_m + \sum_{t=1}^{T}p(t)d(t) + p_m\\
&\leq(1+\lambda) (p_m + \sum_{t=1}^{T}p(t)d(t)) \leq(1+\lambda)\text{OPT},
\end{align*}
where (a) is true from Algorithm \bed. 

(ii) $\hat{\sigma} = \sigma \leq 1$, i.e., $s=1/\lambda$.  With hyperparameter $\lambda$, the algorithm uses the local generator for the first $T^{1/\lambda}$ time slots before switching to the grid, where $T^{1/\lambda} \leq T$.  Then the cost of \bed is  $\text{ALG} = \sum_{t=1}^{T^{1/\lambda}}p_gd(t) + \sum_{t=T^{1/\lambda}+1}^{T}p(t)d(t) + p_m$.  Since $\sigma \leq 1$, the optimal offline solution uses the local generator for the whole duration with cost  $\text{OPT} = \sum_{t=1}^{T}p_gd(t) $.  Then, we have 
\begin{align*}
&\text{ALG}\stackrel{(b)}{\leq} \sum_{t=1}^{T^{1/\lambda}}p_gd(t) + \sum_{t=T^{1/\lambda}+1}^{t = T}p_gd(t) + p_m=\text{OPT}+p_m\\
&\stackrel{(c)}{\leq} \text{OPT} + \lambda (\sum_{t=1}^{T}p_gd(t))\leq \text{OPT} + \lambda \text{OPT}=(1+\lambda)\text{OPT},
\end{align*}
where (b) holds true since $p_g\geq p(t),$ and (c) is true since $T\geq T^{1/\lambda}+1$.
Therefore, \bed is $(1 + {\lambda})$-consistent.  


\begin{algorithm}[!t]
	\caption{\red}
	\begin{algorithmic}
		\label{alg:red}
		\IF{$\hat{\sigma} > 1$}
		\STATE \begin{align*}
		f_1^*(s)=  \begin{cases} 
		\frac{[(1-\lambda)(e-1)+\beta](1-\lambda)}{e-1+\beta}\delta(0), & \text{when }s=-1;\\
		\frac{\lambda e^s}{e-1+\beta}, & \text{when }s \in [0,1]; \\
		\frac{[(1-\lambda)(e-1)+\beta]\lambda}{e-1+\beta}\delta(0), & \text{when }s=\infty;\\
		0, & \text{otherwise}.
		\end{cases}
		\end{align*}
		\ELSE      
		\STATE  \begin{align*}
		f_2^*(s)=  \begin{cases} 
		\frac{\lambda e^s}{e-1+\beta}, & \text{when }s \in [0,1]; \\
		\frac{(1-\lambda)(e -1) + \beta}{e-1+\beta}\delta(0), &\text{when }s = \infty; \\
		0, & \text{otherwise}.
		\end{cases}
		\end{align*}
		\ENDIF
		\STATE Pick a value $s$ randomly according to probably distribution $f_1^*(s)$ or $f_2^*(s)$, and switch to grid electricity starting at the first time $\tau$ where
		$\sum_{t=1}^\tau (p_g - p(t))d(t) \geq  s \cdot p_m. $
	\end{algorithmic}
\end{algorithm}

\subsection{\red: A Randomized Algorithm} 
In randomized algorithms, the decision making is based on random draws from a proper probability distribution function.  First, we emphasize that a randomized algorithm that na\"{\i}vely and based on the guidelines in deterministic algorithm modifies the distribution function~\eqref{eq:dis-zhang} fails to achieve both robustness and consistency at the same time. In particular, a first attempt to change the distribution function is to naturally modify according to the enhancements in deterministic algorithms and obtain the following functions

if $\hat{\sigma} > 1$:
$f_1^*(s)=  \begin{cases} 
\frac{ e^s}{e^{\lambda}-1+\beta}, & s \in [0,\lambda]; \\
\frac{\beta}{e^{\lambda}-1+\beta}\delta(0), & s=\infty;\\
0, & \text{otherwise};
\end{cases}$

if $\hat{\sigma} \leq 1$:
$f_2^*(s)=  \begin{cases} 
\frac{\lambda e^s}{e^{1/\lambda}-1+\beta}, & s \in [0,1/\lambda]; \\
\frac{ \beta}{e^{1/\lambda}-1+\beta}\delta(0), &s = \infty; \\
0, & \text{otherwise}.
\end{cases}$

Our analysis (details in the appendix) demonstrates that with these functions, the randomized algorithm is $\max \Big\{\min \Big\{1/\beta, 1/\lambda\Big\} \cdot \frac{e^{1/\lambda}}{e^{1/\lambda}-1+\beta}, \frac{e^{\lambda}}{e^{\lambda}-1+\beta}\Big\}$-robust and $(1/\beta)$-consistent, hence, with above distribution functions, the consistency could be large as $\beta$ approaches 0.

We develop another randomized algorithm, \red, as summarized in Algorithm~\ref{alg:red}. In \red we modify the probability distribution function of \pred based on $\lambda$ and $\hat{\sigma}$ as in Equation~\eqref{eq:sigma_hat}.  These probability distribution functions are carefully designed such that setting $\lambda$ closer to $0$ raises the density at the optimal predicted value of $s$, while setting $\lambda$ closer to $1$ shifts the distribution towards the original distribution function of \pred.

\begin{theorem}\label{thm:randomized} 
	\emph{\red} achieves a competitive ratio of $\Phi\left[e + \frac{(1-\lambda)(1-\beta)}{\Phi\beta}\right]$, where $\lambda\in(0,1)$, and $\Phi=\frac{1}{e-1+\beta}$. In particular, \emph{\red} is $\left(\Phi\left[e + \frac{(1-\lambda)(1-\beta)}{\Phi\beta}\right]\right)$-robust and $ \left(\Phi\left[e+(\lambda -1)(1-\beta) + \frac{\lambda(1-\lambda)(1-\beta)(e-1)}{\beta }\right] \right)$-consistent.
\end{theorem}

\textbf{Remarks.} (1) Setting $\lambda = 1$ recovers the competitive ratio of $e/(e-1+\beta)$ for the optimal randomized online algorithm~\cite{zhang2018peak}. Further, setting $\lambda =0$ results in a competitive ratio of 1, meaning performing optimally once the learning prediction is accurate.

\textit{Sketch of the proof of theorem~\ref{thm:randomized}:}
Given Proposition~\ref{prop:deterministic}, we compute the expected competitive ratio of \red under several cases.   In below we highlight the competitive ratios in each case, and the detailed derivation of competitive ratios are given in the appendix. 

(i) $\hat{\sigma} > 1,  \sigma \leq 1 $.  Note this is a worst case failed prediction scenario. 
We have 
\begin{align*}
\int_s h(s,\sigma) f_1^*(s) ds = &\Phi [e-(1-\lambda)(1-\beta)\displaybreak[0]
+\frac{(1-\sigma)(1-\lambda)(1-\beta)[(1-\lambda)(e-1)+\beta]}{\sigma}]\displaybreak[1]\\
\leq &\Phi [e +{(1-\lambda)(1-\beta)}/({\Phi\beta})].
\end{align*}

(ii) $\hat{\sigma} > 1,  \sigma > 1 $.  Note this is a best case correct prediction scenario. 
We have 
\begin{align*}
\int_s h(s,\sigma) f_1^*(s)=&\Phi [1/\Phi + \lambda(1-\beta) + \frac{\lambda(1-\lambda)(1-\beta)(\sigma -1)(e-1)}{(\sigma-1)\beta + 1}]\displaybreak[1]\\
\leq&\Phi [1/\Phi   + \lambda(1-\beta) + {\lambda(1-\lambda)(1-\beta)(e-1)}/{\beta }].
\end{align*}
Since this is the best prediction case (prediction error is $0$), we have $\left(\Phi[ e+(\lambda -1)(1-\beta) +  {\lambda(1-\lambda)(1-\beta)(e-1)}/{\beta }]\right)$-consistent.  To prove robustness, we can provide the following upper bound: $\int_s h(s,\sigma) f_1^*(s) ds$
\begin{align*}
&\leq \Phi [e - (1-\lambda)(1-\beta)  +{\lambda(1-\lambda)(1-\beta)(e-1)}/{\beta}]\displaybreak[0]\displaybreak[1]\\
&\leq \Phi[e  +{(1-\lambda)(1-\beta)(e-1+\beta)}/{\beta}]\\
& =\Phi[e  +{(1-\lambda)(1-\beta)}/{(\Phi\beta)}].
\end{align*}

(iii) $\hat{\sigma} \leq 1,   \sigma > 1$.  Note this is a worst case failed prediction scenario.  
We have   $\int_s h(s,\sigma) f_2^*(s)$  
\begin{align*}
= &\Phi \left[ e + \frac{(1-\beta)(1-\lambda)}{(\sigma-1)\beta +1} \left[(\sigma -1)(e-1) - 1  \right]\right] \displaybreak[1]\\
\leq  &\Phi \left[ e + {(1-\beta)(1-\lambda)}/({\Phi\beta }) \right].
\end{align*}
This proves the robustness bound.

(iv) $\hat{\sigma} \leq 1,  \sigma \leq 1 $.  Note this is a best case correct prediction scenario.  
We have   
\begin{align*}
\int_s h(s,\sigma) f_2^*(s) ds 
=\Phi[e -1 + \beta + \lambda(1-\beta)].
\end{align*}

To prove robustness, we have $\int_s h(s,\sigma) f_1^*(s) ds$ \begin{align*}
\leq& \Phi [e - (1-\lambda)(1-\beta) + {(1-\lambda)(1-\beta)}/({\Phi\beta})]\displaybreak[0]\\
\leq &\Phi [e + {(1-\lambda)(1-\beta)}/({\Phi\beta})].
\end{align*}

Putting together the results above, we obtain the robustness and consistency given in Theorem~\ref{thm:randomized}. 

\subsubsection{Extending the algorithms to the general case of non-negative demand:}
The above competitive ratios can be extended to the general problem of non-negative integer demand.  This is done by dividing the integer demand $d(t)$ into multiple subproblem layers with 0 or 1 demand.  At a given layer $i$, the layered demand at time $t$ is 1 if $d(t) \leq i$ and 0 otherwise.  Then the result in ~\cite[Theorem~3]{zhang2018peak} can be applied. The competitive ratio of an algorithm which solves the subproblem with 0/1 demand is an upper bound to the competitive ratio of an algorithm which solves the general  integer demand problem using the layering strategy.

\section{Experiments}
\label{sec:exp}


\subsubsection{Overview of data traces:} 
\begin{figure*}
	\center
	\begin{minipage}[b]{.32\textwidth}
		\includegraphics[width=0.83\textwidth]{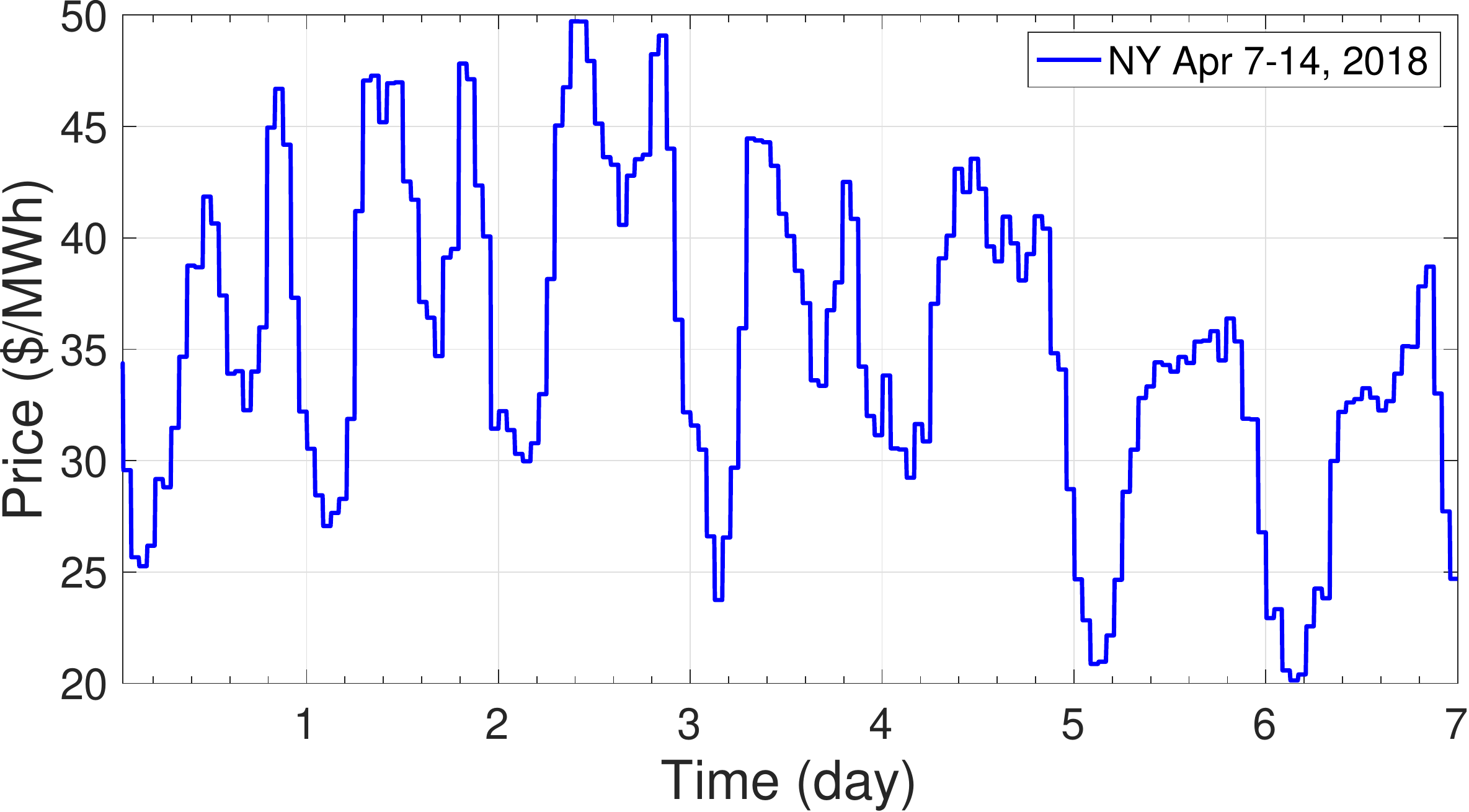}		
		\caption{Time-varying energy prices for New York electricity market}
		\label{fig:ny_prices}%
	\end{minipage}
	\hspace{2mm}
	\begin{minipage}[b]{.32\textwidth}
		\includegraphics[width=0.83\textwidth]{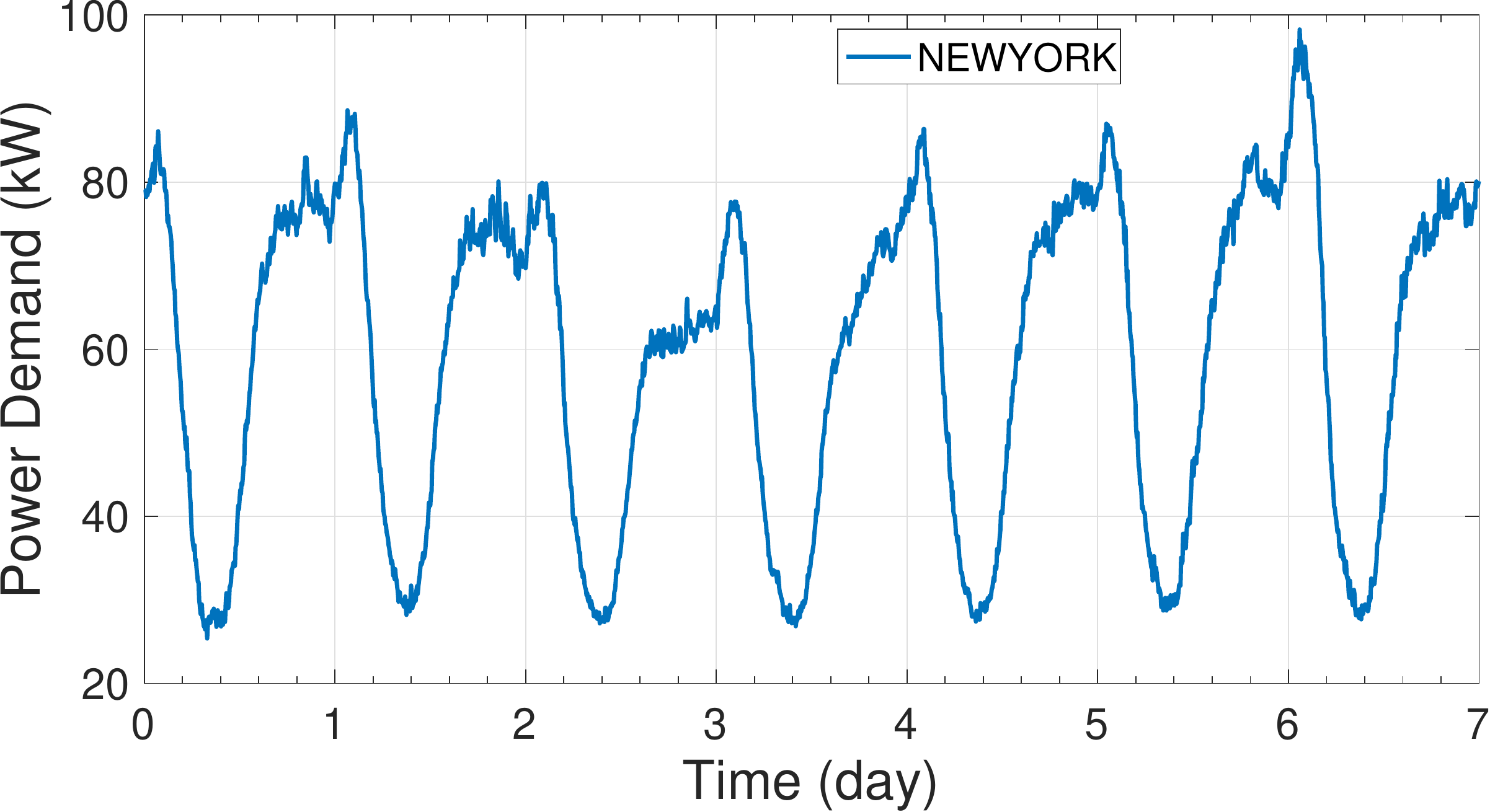}	
		\caption{Time-varying energy demands for an Akamai data center in New York}
		\label{fig:us_city}%
	\end{minipage}
	\begin{minipage}[b]{.32\textwidth}
		\includegraphics[width=1\textwidth]{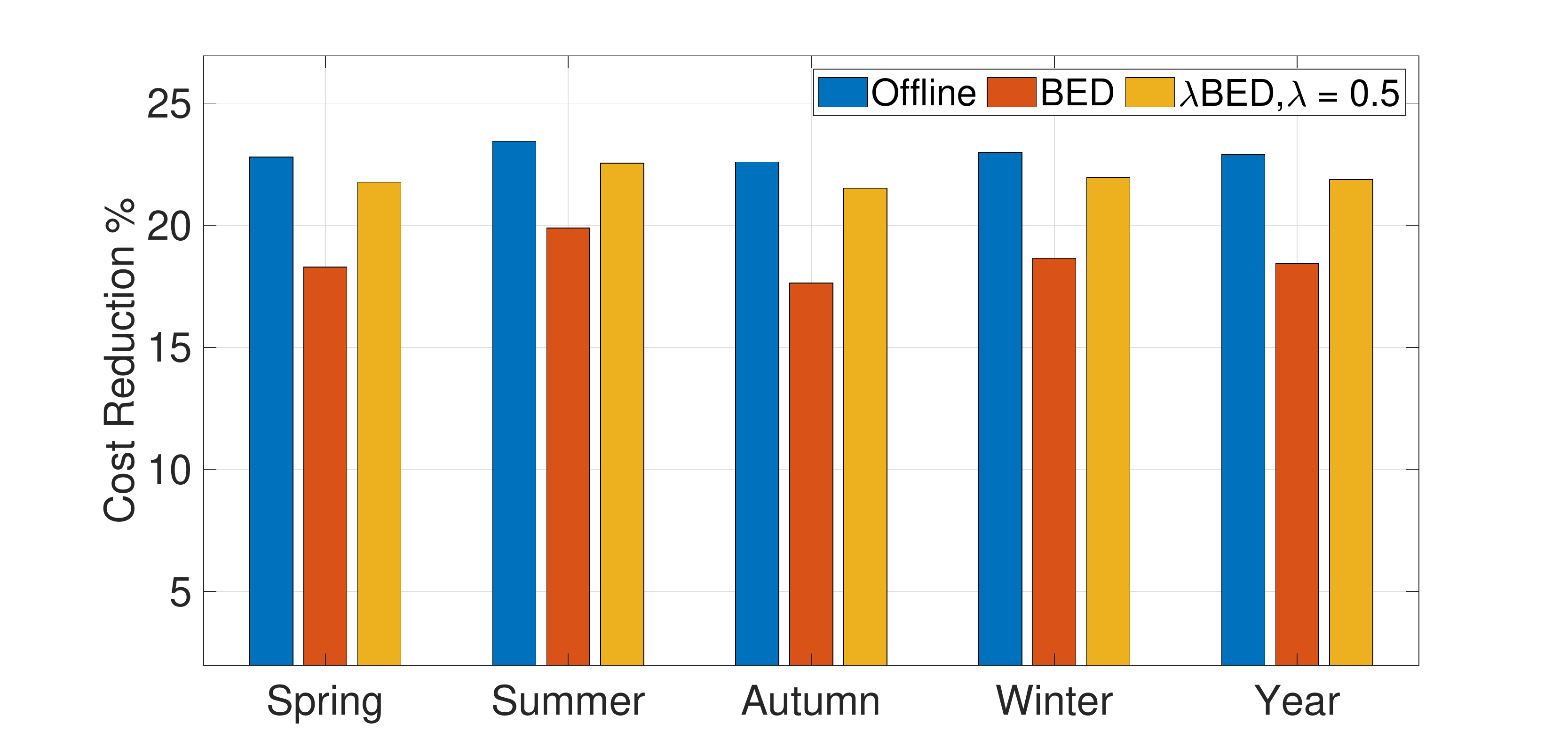}	\vspace{-2mm}
		\caption{Cost reduction of different algorithms for different seasons and the year}
		\label{fig:season}
	\end{minipage}
\end{figure*}

We use spot energy prices for 2018 from New York Electricity Market (NYISO). The values of spot prices changes in hourly manner. For example, the spot prices in April 2018 changes between \$13.69/MWh and \$64.62/MWh. A snapshot of energy prices for one week in April 2018 shown in Figure~\ref{fig:ny_prices}, demonstrates that the spot prices change in irregular patterns. We use energy demands for an Akamai data center in New York~\cite{nygren2010akamai}. A sample one week trajectory of energy demands is depicted in Figure~\ref{fig:us_city}. The diurnal pattern of energy demand makes it possible to predict these values using machine learning models, motivating the proposed learning-assisted algorithms in this paper. 

\subsubsection{Settings and comparison algorithms:} Unless otherwise mentioned, we use the following values for parameters. By setting the length of each slot to one hour, we set $T = 24 \times 30$ that represents the billing cycle of one month, which is common for the electricity bill. The value of $p_m$ is set to be roughly $~100 \times \max_{t\in\mathcal{T}} p(t)$, which is based on the common practice in the U.S. utilities, such as PG\&E and Duke Energy. The cost of local generator is set to $p_g = \max_{t\in\mathcal{T}}p(t)$. Finally, the capacity of the local generator is set to be roughly 60\% of the energy demand.
In experiments, we report the cost reduction of different algorithms as compared to a benchmark of not using local generators. Also, we report the empirical competitive ratios of online algorithms that simply shows the ratio between the cost of an online algorithm and offline optimum. Last, we compare the cost of our proposed learning-assisted algorithms \bed to the pure online algorithm \pbed~\cite{zhang2018peak}. 

\begin{figure*}[!h]
	\center
	\begin{minipage}[b]{.33\textwidth}
		\includegraphics[width=0.92\textwidth]{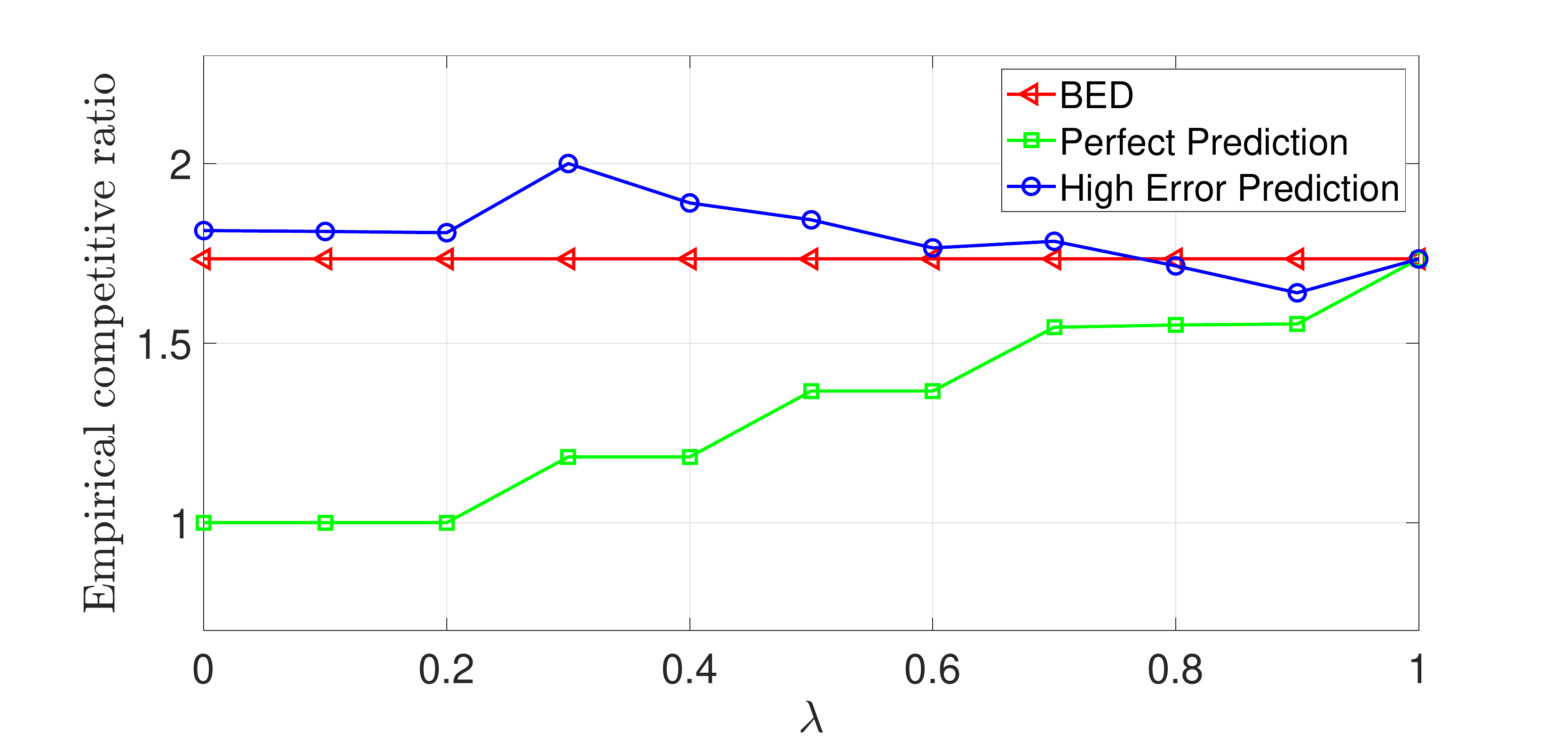}	
		\caption{Hyperparameter $\lambda$}
		\label{fig:lambda}%
	\end{minipage}
	\begin{minipage}[b]{.33\textwidth}
		\includegraphics[width=0.83\textwidth]{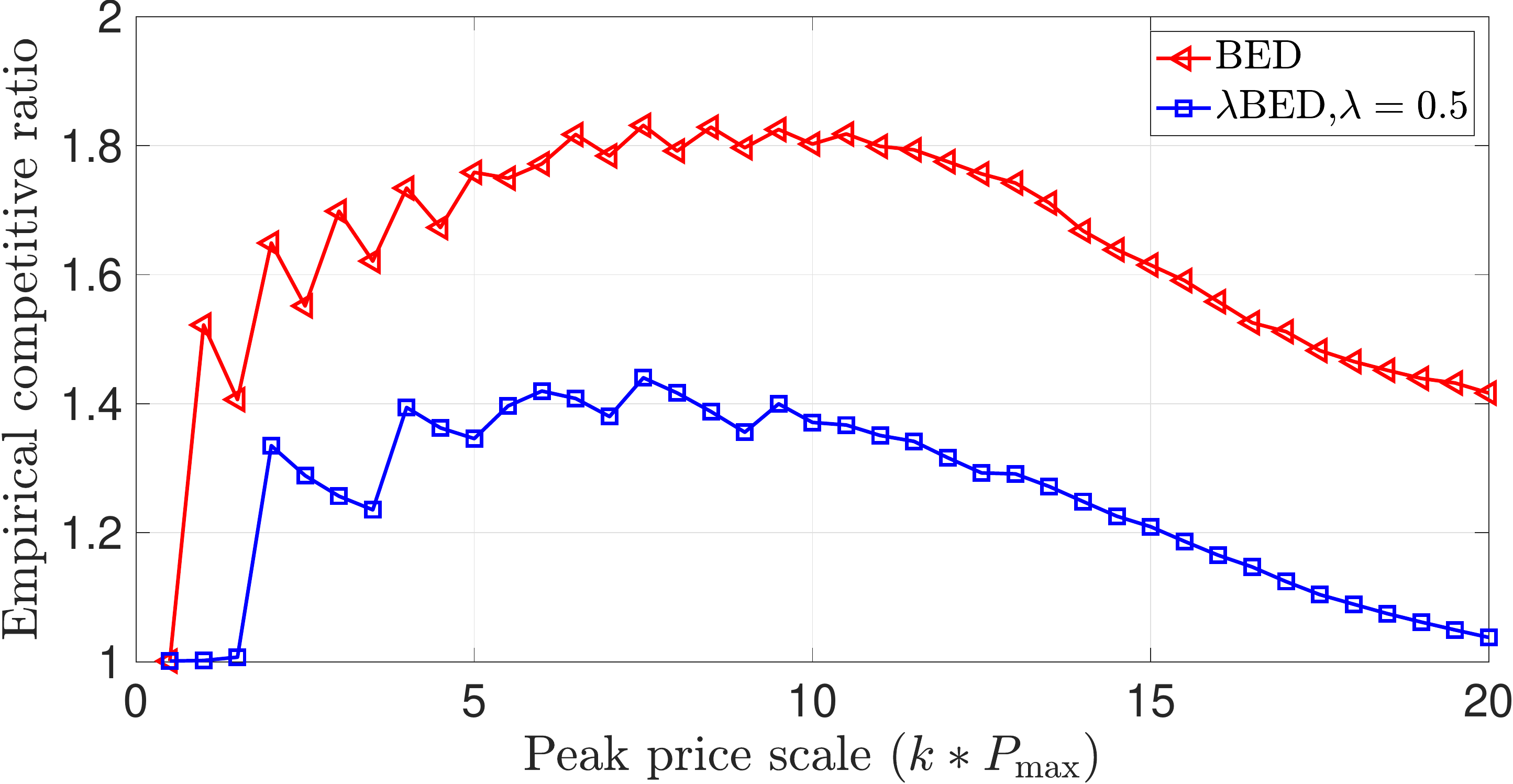}	
		\caption{Peak price}
		\label{fig:peak}
	\end{minipage}
	\begin{minipage}[b]{.33\textwidth}
		\includegraphics[width=0.92\textwidth]{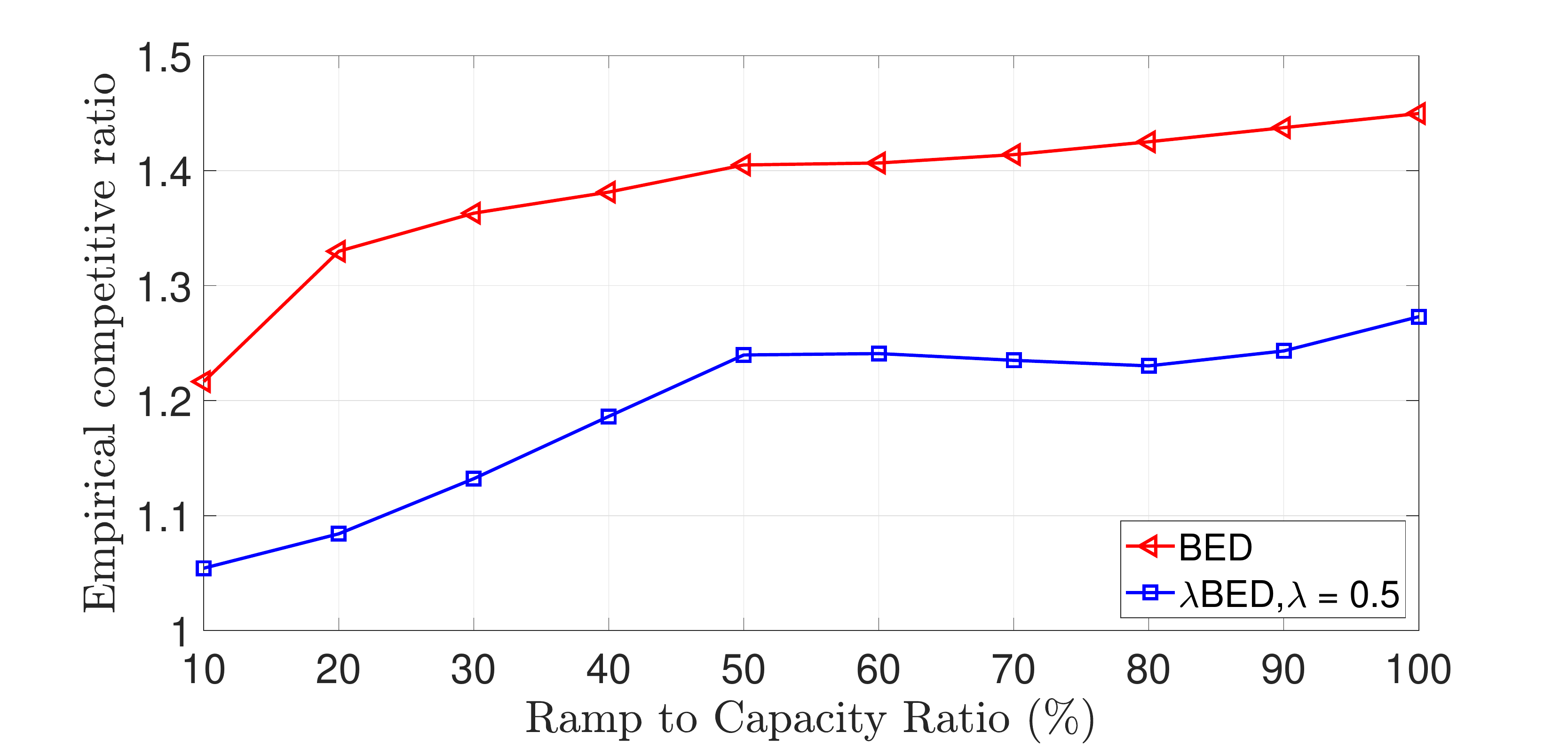}	
		\caption{Ramp constraint}
		\label{fig:ramp}
	\end{minipage}	
\end{figure*}

\begin{figure}[t]
	\centering
	\subfigure[Competitive ratio]{	\vspace{0pt}		\includegraphics[width=0.25\textwidth]{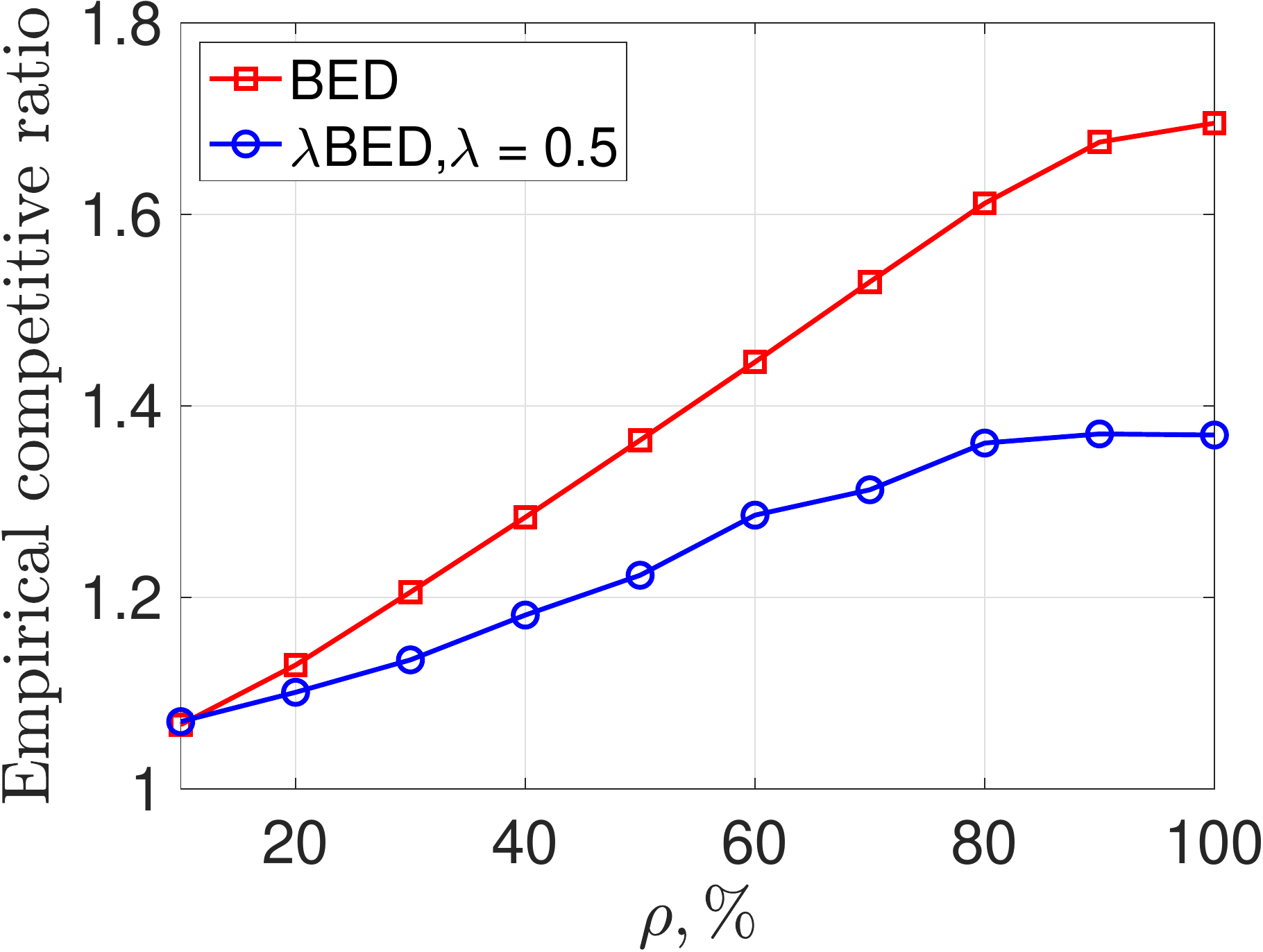}		
		\label{fig:capacity}}
	\subfigure[Cost reduction]{
		\vspace{0pt}		\includegraphics[width=0.25\textwidth]{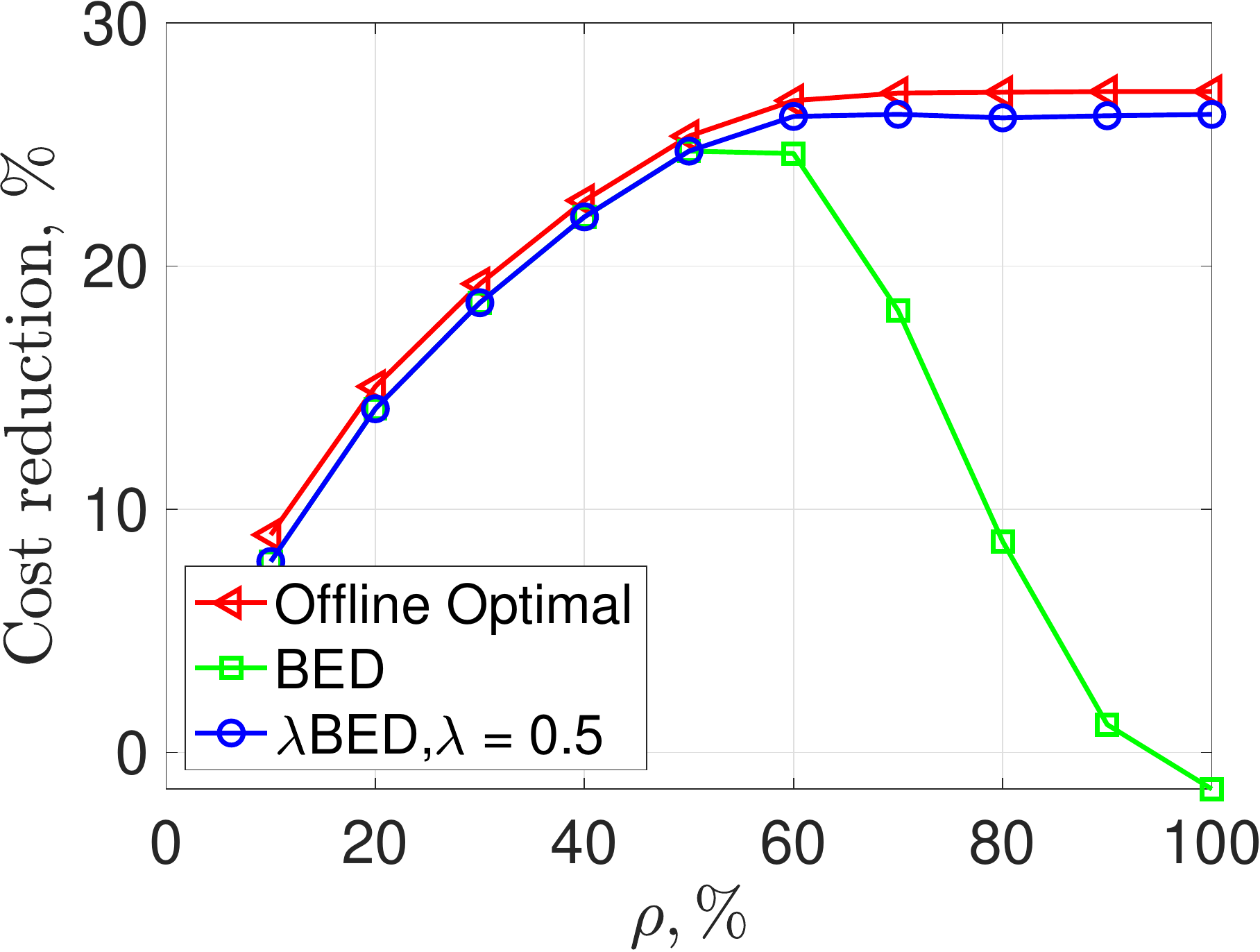}		
		\label{fig:capacity_cost_reduction}}
	\caption{Generation capacity}
	\label{fig:gen_cap}
\end{figure}



\subsubsection{Seasonal benefits of employing local generators:}
In this experiment, we report the cost reduction of optimal offline, \pbed, and \bed with $\lambda=0.5$ for different seasons, and for the year. For \bed, we assume that there is a prediction for the demand and price values. The main goal is to show the performance of \bed with some prediction, so, we use a simple model as the predicted values. We calculate the prediction values as follows. For each time $t$, the predicted grid price $\hat{p}(t)$ and predicted demand $\hat{d}(t)$ are simulated as $\hat{p}(t) = p(t) + \epsilon_1$ and $\hat{d}(t) = d(t) + \epsilon_2$. The error terms $\epsilon_1,\epsilon_2$ are drawn from normal distributions with mean 0 and standard deviations $\sigma_1$ and $\sigma_2$. The values of $\sigma_1$ and $\sigma_2$ are set to be half of maximum spot price and energy demand, respectively. From the 68-95-99 rule of normal distributions,the generated error that is incorrect by more than half the maximum true demand or price for about 34\% of time steps. Figure~\ref{fig:season} show that (i) the average cost reduction of \bed over the year is within $4.44\%$ of the best possible cost reduction in offline; and (ii) \bed outperforms \pbed by $15.13\%$, which means that by adding some noisy prediction, one can substantially improve the performance of online algorithms in practice. 

\subsubsection{The impact of hyperparameter $\lambda$:} Introducing hyperparameter in algorithm design allows effective usage of predictions in algorithmic actions. Specifically, setting $\lambda$ close to 0, represents more trust on prediction, and $\lambda$ close to 1, represents almost no trust on prediction. To scrutinize the impact of $\lambda$ on the performance of \bed, in Figure~\ref{fig:lambda}, we vary the value of $\lambda$ from 0 to 1.  We report the competitive ratio of \bed for one single problem instance, however, with feeding \bed with a perfect prediction and an extremely erroneous prediction. 
The notable observations are as follow: (1) With perfect prediction, and $\lambda \leq 0.2$, \bed achieves the optimal performance, i.e., competitive ratio of 1. (2) With high prediction error, improper values of $\lambda$ that leads to high trust on prediction, leads to even worse performance than the pure online algorithm. (3) However, with proper setting of $\lambda \geq 0.9$, \bed achieves better performance that \pbed even with extremely erroneous prediction. This experiment signifies the importance of setting right values for the hyperparameter. \textit{More importantly, the obtained theoretical and empirical results demonstrate that learning-assisted algorithm does not guarantee better performance regardless of the value of hyperparameter. Instead, it guarantees that it is always possible to find the right value of hyperparameter such that the performance of learning-assisted algorithm is better than pure online one.}

\subsubsection{The impact of peak price:} The peak price $p_m$ is an important parameter that can make impact on the contribution of peak charge and also on the break-even point $\sigma$ as defined in Equation~\eqref{eq:sigma}. In a case study, it has been shown that the peak charge varies substantially in different geographical locations, e.g., from 20\% to 80\% of the total electricity bill~\cite{xu2013reducing}. In this experiment, we investigate the impact of this parameter on different algorithms. We scale the value of peak price from $1\times$ to $20\times$ of its original value and report the corresponding competitive ratio values in Figure~\ref{fig:peak}.
The result shows that the competitive ratio of \bed is constantly better than \pbed. More interestingly, the result shows that the cost ratios are better in two extremes of low and high peak prices. This is reasonable since with low peak prices it makes sense to frequently use the electric grid, since the peak charge is not significant. On the other hand, with high peak prices, it is an optimal decision to fully utilize the local generation, and then satisfy the residual from the grid. So, despite the uncertainty of the input, the decisions in these two extreme regimes are pretty straightforward.  The problem is more intriguing once the peak price is neither too low nor too high. In these regions, \bed outperforms \pbed substantially.


\subsubsection{The impact of ramp constraint:} The algorithms proposed in this paper work for fast-response generators. In practice, there are several generators that are slow-response---they cannot switch their output level quickly. The proposed algorithms could be straightforwardly modified to incorporate ramp constraints. Specifically, let $R$ be the ramp constraints, hence we have $|u(t)-u(t-1) \leq R, \forall t$, i.e., the changes in generator output level should be always less than $R$. We can easily modify \bed and \pbed, as explained in~\cite[Section 4]{zhang2015peak}, to reflect the ramp constraint. The idea is to first run the algorithm without the ramp constraints, and then, project the obtained values to the feasible region to respect the ramp constraints. To show the impact of ramp constraints, in Figure~\ref{fig:ramp}, we vary the ramp to capacity ratio from 10\% to 100\%, and report the empirical competitive ratios of \pbed and \bed. The result shows that \bed always achieves better performance than \pbed. More interestingly, the competitive ratios increase once we relax the ramp constraints. This is because with ramp constraints, the feasible region of the optimization problem is restricted, so, the optimal offline has limited flexibility to optimize the cost, and by relaxing this, the optimal offline is more powerful and online algorithm fails to achieve a close-to-optimal performance.

\subsubsection{The impact of local generation capacity:}
A  drawback of pure online algorithms such as \pbed is that they are too conservative to make a decision. Specifically, \pbed waits long to see the break-even point and switch to the grid. This might degrade the performance of the algorithm substantially. An example of such performance degradation is once the capacity of the generator is above 60\% of the total energy demand (see Figure 5 in~\cite{zhang2018peak}). 
By leveraging ML prediction in \bed, however, we can effectively prevent this performance degradation. To show this, in this experiment, we investigate the cost saving of different algorithms as the capacity of generator changes. Toward this, we define $\rho = C/\max_t d(t)$ as the ratio between the capacity of generator and the maximum energy demand, and change this value from 0.1 to 1. The results are shown in Figure~\ref{fig:gen_cap}, where Figure~\ref{fig:capacity} shows the competitive ratios of \pbed and \bed. To highlight the benefit of learning-assisted competitive algorithms in resolving the pessimistic decision making of pure online algorithm, in Figure~\ref{fig:capacity_cost_reduction}, we report the cost reductions as compared to a baseline without local generation. Notable observations are as follow. First, with $\rho \leq 50\%$, both \bed performs slightly better than \pbed. However, with $\rho > 60\%$ the performance of \pbed degrades substantially, while the cost reduction of \bed as compared to the optimal offline degrades slightly. \textit{We consider this observation as the most important motivation to use learning-assisted online algorithms for tackling online problems.}

\section{Concluding Remarks and Future Directions}
In this paper, we developed learning-assisted online algorithms for peak-aware energy scheduling problem. The proposed algorithms are provably robust against poor prediction of the machine-learning and behave optimally if the machine learning prediction is accurate. Experimental results using real data traces verify the theoretical results and demonstrate the superiority of the proposed algorithms as compared to existing pure online algorithms. 
As for future directions, we plan to extend the results for more complicated problem setting by considering energy storage as generators. In solution design, we plan to characterize the competitive ratio as a function of error of ML-predictor. 
\begin{acks}
This work is supported by a Google Faculty Research Award and NSF grant CNS-1908298.
\end{acks}

\bibliographystyle{ACM-Reference-Format}
\bibliography{paper.draft.bib}

\appendix

\section{Detailed Proof of Theorem 1}
\label{app:deteministic}
We first consider the robustness. 
The worst-case cost ratio for a general deterministic algorithm $\mathcal{A}_s$ with parameter $s$ is when $\sigma = s$, where the online algorithm pays for the peak charge premium but has no net demand to serve anymore.  From Proposition 1, this worst case cost ratio $\max_{\sigma}h(\mathcal{A}_s,\sigma)$ is 
\begin{equation}
\label{eq:worst-ratio_1}
\max_{\sigma}h(\mathcal{A}_s,\sigma) = \begin{cases}
1 + \frac{1}{s}(1-\beta), & \text{if }s\leq 1; \\
1 + \frac{s(1-\beta)}{(s-1)\beta + 1}, & \text{otherwise}. 
\end{cases}
\end{equation}

We compute the competitive ratio of  \bed under two cases

(i) $\hat{\sigma} >1$: According to \bed, $s = \lambda < 1$. 
From~(\ref{eq:worst-ratio_1}), we have $\texttt{CR}(\mathcal{A}_{\lambda}) = 1 + (1-\beta)/\lambda.$

(ii) When $\hat{\sigma} \leq 1$: According to \bed, $s = 1/\lambda > 1$. 
From~(\ref{eq:worst-ratio_1}), we have $\texttt{CR}(\mathcal{A}_{1/\lambda}) = 1 + \frac{(1-\beta)/\lambda}{(1/\lambda-1)\beta + 1}.$


This means that  \bed is $(1 + (1-\beta)/\lambda)$-robust.  Note that setting $\lambda = 1$ recovers the competitive ratio of the optimal online algorithm.

Next, we consider the consistency.    For consistency guarantees, we compute the competitive ratio assuming the predictions are correct.  There are two cases to consider here

(i)  $\hat{\sigma} = \sigma >1$.  
With hyperparameter $\lambda$, the algorithm uses the local generator for the first $T^{\lambda}$ time slots before switching to the grid.  Then the cost of the algorithm is  $\texttt{ALG}= \sum_{t=1}^{T^{\lambda}}p_gd(t) + \sum_{t=T^{\lambda}+1}^{T}p(t)d(t) + p_m$.  Since $\sigma > 1$, the optimal offline solution uses the grid for the whole duration with cost  $\texttt{OPT}= \sum_{t=1}^{T}p(t)d(t) + p_m$.  Then we have

\begin{align*}
\texttt{ALG} &=\sum_{t=1}^{T^{\lambda}}p_gd(t) + \sum_{t=T^{\lambda}+1}^{ T}p(t)d(t) + p_m\\
&=\sum_{t=1}^{T^{\lambda}}(p_g - p(t))d(t) + \sum_{t=1}^{T}p(t)d(t) + p_m\\
&\stackrel{(a)}{\leq}\lambda \cdot p_m + \sum_{t=1}^{T}p(t)d(t) + p_m\\
&=(1+\lambda) \cdot p_m + \sum_{t=1}^{T}p(t)d(t) \\
&\leq(1+\lambda) (p_m + \sum_{t=1}^{T}p(t)d(t))\\ 
&\leq(1+\lambda)\texttt{OPT},
\end{align*}
where (a) is true from Algorithm \bed. 

(ii) $\hat{\sigma} = \sigma \leq 1$.  
With hyperparameter $\lambda$, the algorithm uses the local generator for the first $T^{1/{\lambda}}$ time slots before switching to the grid, where $T^{1/{\lambda}} \leq T$.  Then the cost of the algorithm is  $\texttt{ALG}= \sum_{t=1}^{T^{1/{\lambda}}}p_gd(t) + \sum_{t=T^{1/{\lambda}}+1}^{T}p(t)d(t) + p_m$.  Since $\sigma \leq 1$, the optimal offline solution uses the local generator for the whole duration with cost  $\texttt{OPT}= \sum_{t=1}^{T}p_gd(t) $.  Then we have 
\begin{align*}
\texttt{ALG} &= \sum_{t=1}^{T^{1/{\lambda}}}p_gd(t) + \sum_{t=T^{1/{\lambda}}+1}^{T}p(t)d(t) + p_m\\
& \leq \sum_{t=1}^{T^{1/{\lambda}}}p_gd(t) + \sum_{t=T^{1/{\lambda}}+1}^{T}p_gd(t) + p_m\displaybreak[0]\\
&= \texttt{OPT} + p_m\\
&\leq \texttt{OPT} + \lambda \Bigg(\sum_{t=1}^{T^{1/{\lambda}}+1}(p_g - p(t))d(t)\Bigg)\displaybreak[1]\\
&\stackrel{(b)}{\leq} \texttt{OPT} + \lambda \Bigg(\sum_{t=1}^{T}p_gd(t)\Bigg)\displaybreak[2]\\
&\leq \texttt{OPT} + \lambda \texttt{OPT}\\
&\leq (1+\lambda)\texttt{OPT}, 
\end{align*}

where (b) is true since $T\geq T^{1/{\lambda}}+1$ and $p(t)\geq 0.$

This means that \bed is $(1 + {\lambda})$-consistent.  Note that setting $\lambda = 0$ results in a competitive ratio of $1$, which means doing optimally once learning prediction is accurate. 

\section{Detailed Proof of Theorem 2}
(i) $\hat{\sigma} > 1,  \sigma \leq 1 $.  Note this is a worst case failed prediction scenario. 
\begin{align*}
&\int_s h(s,\sigma) f_1^*(s) ds \\
=& \int_0^{\sigma} \Bigg[  1 + \frac{1-\sigma+s}{\sigma}(1-\beta)\Bigg]\frac{\lambda e^s}{e-1+\beta} ds
+ \Bigg[1 + \frac{1-\sigma -1}{\sigma}(1-\beta)\Bigg]\frac{[(1-\lambda)(e-1)+\beta](1-\lambda)}{e-1+\beta} \\
& +\int_{\sigma}^{1}\frac{\lambda e^s}{e-1+\beta} ds + \frac{[(1-\lambda)(e -1) + \beta]\lambda}{e-1+\beta}\\
=& \frac{1}{e-1+\beta} \Bigg[\int_0^{1}\lambda e^s ds + \lambda(1-\beta)\int_0^{\sigma} \frac{1-\sigma+s}{\sigma} e^s ds
+ (1-\lambda)(e-1)+\beta \\
& + \frac{-\sigma}{\sigma}(1-\beta)[(1-\lambda)(e-1)+\beta](1-\lambda)\Bigg]
= \frac{1}{e-1+\beta} \Bigg[\lambda (e-1)  + (1-\lambda)(e-1)+\beta \\
& + \lambda(1-\beta) \frac{(1-\sigma)(e^\sigma-1)+e^\sigma(\sigma-1)+1}{\sigma} 
- (1-\beta)[(1-\lambda)(e-1)+\beta](1-\lambda)\Bigg]\\
= &\frac{1}{e-1+\beta} \Bigg[e-1  + \beta + \lambda(1-\beta) \frac{\sigma}{\sigma} -(1-\beta)[(1-\lambda)(e-1)+\beta](1-\lambda)\Bigg]\displaybreak[1]\\
= &\frac{1}{e-1+\beta} \Bigg[e-(1-\lambda)(1-\beta) -(1-\sigma)(1-\lambda)(1-\beta)[(1-\lambda)(e-1)+\beta]\Bigg]\displaybreak[1]\\
\leq &\frac{1}{e-1+\beta} \Bigg[e +\frac{(1-\lambda)(1-\beta)(e-1+\beta)}{\beta}\Bigg].
\end{align*}
This proves the robustness bound.  If $\lambda = 1$, we have $\left(\frac{e}{e-1+\beta}\right)$-robust.

(ii) $\hat{\sigma} > 1,  \sigma > 1 $.  Note this is a best case correct prediction scenario. 

\begin{align*}
&\int_s h(s,\sigma) f_1^*(s) ds\\ 
=&\int_0^1 \Bigg[ 1+\frac{s(1-\beta)}{(\sigma-1)\beta + 1}\Bigg] \frac{\lambda e^s}{e-1+\beta}ds
+ \Bigg[ 1+\frac{(-1)(1-\beta)}{(\sigma-1)\beta + 1}\Bigg]\frac{[(1-\lambda)(e-1)+\beta](1-\lambda)}{e-1+\beta}\\
&+ \Bigg[ 1+\frac{(\sigma-1)(1-\beta)}{(\sigma-1)\beta + 1}\Bigg]\frac{[(1-\lambda)(e-1)+\beta]\lambda}{e-1+\beta}\\
\leq&\frac{1}{e-1+\beta} \Bigg[\int_0^1 \lambda e^s ds +  \frac{1-\beta}{(\sigma-1)\beta + 1}\int_0^1 \lambda se^s ds - 0
+ [(1-\lambda)(e-1)+\beta] 
+\frac{(\sigma -1)(1-\beta)}{(\sigma-1)\beta + 1} \cdot [(1-\lambda)(e-1)+\beta]\lambda\Bigg]\\
=&\frac{1}{e-1+\beta} \Bigg[e-1 + \beta +  \frac{\lambda(1-\beta)}{(\sigma-1)\beta + 1}\Bigg[1  +  (\sigma -1)
\cdot[(1-\lambda)(e-1)+\beta]\Bigg]\Bigg]\displaybreak[0]\\
=&\frac{1}{e-1+\beta} \Bigg[e-1 + \beta +  \frac{\lambda(1-\beta)}{(\sigma-1)\beta + 1}\Bigg[1  +  (\sigma -1)\beta
+ (\sigma -1)(1-\lambda)(e-1)\Bigg]\Bigg]\displaybreak[1]\\
=&\frac{1}{e-1+\beta} \Bigg[e-1 + \beta + \lambda(1-\beta)
+ \frac{\lambda(1-\lambda)(1-\beta)(\sigma -1)(e-1)}{(\sigma-1)\beta + 1}\Bigg]\displaybreak[2]\\
\leq&\frac{1}{e-1+\beta} \Bigg[e-1 + \beta + \lambda(1-\beta)
+ \frac{\lambda(1-\lambda)(1-\beta)(\sigma -1)(e-1)}{(\sigma-1)\beta }\Bigg]\displaybreak[3]\\
=&\frac{1}{e-1+\beta} \Bigg[e-1 + \beta + \lambda(1-\beta) + \frac{\lambda(1-\lambda)(1-\beta)(e-1)}{\beta }\Bigg].
\end{align*}

If $\lambda = 1$, we have $\left(\frac{e}{e-1+\beta}\right)$-robust.  Since this is the best prediction case, we have $\left(\frac{1}{e-1+\beta}[ e+(\lambda -1)(1-\beta) +  {\lambda(1-\lambda)(1-\beta)(e-1)}/{\beta }]\right)$-consistent.  To prove robustness, we can provide the following upper bound:
\begin{align*}
&\int_s h(s,\sigma) f_1^*(s) ds\displaybreak[0]\\
&\leq \frac{1}{e-1+\beta} \Bigg[e - (1-\lambda)(1-\beta)   +\frac{\lambda(1-\lambda)(1-\beta)(e-1)}{\beta}\Bigg]\displaybreak[0]\displaybreak[1]\\
&\leq \frac{1}{e-1+\beta} \Bigg[e   +\frac{(1-\lambda)(1-\beta)(e-1)}{\beta}\Bigg]\displaybreak[1]\\
&\leq \frac{1}{e-1+\beta} \Bigg[e  +\frac{(1-\lambda)(1-\beta)(e-1+\beta)}{\beta}\Bigg].
\end{align*}

(iii) $\hat{\sigma} \leq 1,   \sigma > 1$.  Note this is a worst case failed prediction scenario.  

\begin{align*}
& \int_s h(s,\sigma) f_2^*(s) ds\\ &= \int_0^1 \left[  1 + \frac{s(1-\beta)}{(\sigma-1)\beta +1}\right]\frac{\lambda e^s}{e-1+\beta} ds \\
&+ \left[1 + \frac{(\sigma-1)(1-\beta)}{(\sigma-1)\beta +1}\right]\frac{(1-\lambda)(e-1)+\beta}{e-1+\beta} \nonumber\displaybreak[0]\\
=& \frac{1}{e-1+\beta} \left[ \int_0^{1} \lambda e^s ds + (1-\lambda)(e-1) + \beta
+\frac{(1-\beta)}{(\sigma-1)\beta +1} \left[\int_0^{1} \lambda se^s ds  + (\sigma -1)((1-\lambda)(e-1)+\beta)  \right]\right] \nonumber\displaybreak[1]\\
= &\frac{1}{e-1+\beta} \left[ e-1 + \beta +\frac{(1-\beta)}{(\sigma-1)\beta +1} \left[\lambda  + (\sigma -1)((1-\lambda)(e-1)+\beta)  \right]\right] \nonumber\displaybreak[2]\\
=& \frac{1}{e-1+\beta} \left[ e-1 + \beta+\frac{(1-\beta)}{(\sigma-1)\beta +1} \left[(\sigma -1)\beta
+ 1 + (\lambda-1)  + (\sigma -1)(1-\lambda)(e-1)  \right]\right] \nonumber\displaybreak[3]\\
= &\frac{1}{e-1+\beta} \left[ e-1 + \beta + 1 - \beta +\frac{(1-\beta)}{(\sigma-1)\beta +1} \left[(\lambda-1)  + (\sigma -1)(1-\lambda)(e-1)  \right]\right] \\
=& \frac{1}{e-1+\beta} \left[ e + \frac{(1-\beta)(1-\lambda)}{(\sigma-1)\beta +1} \left[(\sigma -1)(e-1) - 1  \right]\right] \\
\leq&  \frac{1}{e-1+\beta} \left[ e + \frac{(1-\beta)(1-\lambda)}{(\sigma-1)\beta } \left[(\sigma -1)(e-1) \right]\right] \\\displaybreak
= & \frac{1}{e-1+\beta} \left[e + \frac{(1-\beta)(1-\lambda)(e-1)}{\beta } \right] \\
= & \frac{1}{e-1+\beta} \left[ e + \frac{(1-\beta)(1-\lambda)(e-1+\beta)}{\beta } \right]. 
\end{align*}
This proves the robustness bound.

(iv) $\hat{\sigma} \leq 1,  \sigma \leq 1 $.  Note this is a best case correct prediction scenario.  
\begin{align*}
&\int_s h(s,\sigma) f_2^*(s) ds \\
=& \int_0^{\sigma} \Bigg[  1 + \frac{1-\sigma+s}{\sigma}(1-\beta)\Bigg]\frac{\lambda e^s}{e-1+\beta} ds \displaybreak[0]
+ \int_{\sigma}^{1}\frac{\lambda e^s}{e-1+\beta} ds + \frac{(1-\lambda)(e -1) + \beta}{e-1+\beta}\displaybreak[1]\\
=&\frac{1}{e-1+\beta}\Bigg[\int_0^1 \lambda e^s ds + (1-\lambda)(e -1) + \beta 
+ \lambda(1-\beta) \int_{0}^{\sigma} \frac{1-\sigma+s}{\sigma}e^s ds\Bigg]\displaybreak[1]\\
=&\frac{1}{e-1+\beta}\Bigg[\lambda (e-1) + (1-\lambda)(e -1) + \beta 
+ \lambda(1-\beta) \frac{(1-\sigma)(e^{\sigma}-1)+e^{\sigma}(\sigma-1)+1}{\sigma}\Bigg]\displaybreak[1]\\
=&\frac{1}{e-1+\beta}\Bigg[e -1 + \beta 
+ \lambda(1-\beta)\frac{\sigma}{\sigma} \Bigg]\displaybreak[1]\\
=&\frac{1}{e-1+\beta}\Bigg[e -1 + \beta + \lambda(1-\beta) \Bigg].
\end{align*}
For $\lambda = 1$, it is $\left(\frac{e}{e-1+\beta}\right)$-robust.   If $\lambda = 0$, i.e., prediction error is $0$, it is $1$-consistent. 

To prove robustness, we have
\begin{align*}
&\int_s h(s,\sigma) f_1^*(s) ds
\leq \frac{1}{e-1+\beta} \Bigg[e - (1-\lambda)(1-\beta)\Bigg]\displaybreak[0]\\
\leq& \frac{1}{e-1+\beta} \Bigg[e - (1-\lambda)(1-\beta) + \frac{(1-\lambda)(1-\beta)(e-1+\beta)}{\beta}\Bigg]\displaybreak[1]\\
\leq &\frac{1}{e-1+\beta} \Bigg[e + \frac{(1-\lambda)(1-\beta)(e-1+\beta)}{\beta}\Bigg].
\end{align*}

This means that \red is $\left(\frac{1}{e-1+\beta}[e + {(1-\lambda)(1-\beta)(e-1+\beta)}/{\beta}]\right)$-robust and $\left(\frac{1}{e-1+\beta}[ e+(\lambda -1)(1-\beta) + {\lambda(1-\lambda)(1-\beta)(e-1)}/{\beta }]\right) $-consistent. Note that setting $\lambda = 1$ for robustness recovers the competitive ratio of $(\frac{e}{e-1+\beta})$ from the optimal randomized online algorithm, and setting $\lambda = 0$ for consistency recovers the competitive ratio of $1$.

\section{The Robustness and Consistency Analysis of \red with Direct Extension of \pred}

We note that the randomized algorithm that naively modifies the distribution function proposed in equation (2) fails to achieve both robustness and consistency at the same time. In particular, a first attempt to change the distribution function is to naturally modify according to the enhancements in deterministic algorithms and obtain the following functions: 

if $\hat{\sigma} > 1$:
$$f_1^*(s)=  \begin{cases} 
\frac{ e^s}{e^{\lambda}-1+\beta}, & s \in [0,\lambda]; \\
\frac{[\beta]\lambda}{e^{\lambda}-1+\beta}\delta(0) & s=\infty;\\
0, & \text{o.w.},
\end{cases}$$
if $\hat{\sigma} \leq 1$:
$$f_1^*(s)=  \begin{cases} 
\frac{\lambda e^s}{e^{1/\lambda}-1+\beta}, & s \in [0,1/\lambda]; \\
\frac{ \beta}{e^{1/\lambda}-1+\beta}\delta(0), &s = \infty; \\
0, & \text{o.w.}
\end{cases}$$

Our analysis in below demonstrates that with these functions \red is $\max \Big\{\min \Big\{1/\beta, 1/\lambda\Big\} \cdot \frac{e^{1/\lambda}}{e^{1/\lambda}-1+\beta}, \frac{e^{\lambda}}{e^{\lambda}-1+\beta}\Big\}$-robust and $(1/\beta)$-consistent. This means that with above distribution functions the consistency could be large as $\beta$ approaches 0. 

(i) $\hat{\sigma} >1, \sigma \leq \lambda < 1$.  Note this is a worst case failed prediction scenario.  
\begin{align*}
&\int_s h(s,\sigma) f_1^*(s) ds \\
&= \int_0^{\sigma} \Bigg[ 1 + \frac{1-\sigma + s}{\sigma} (1-\beta)\Bigg]\frac{e^s}{e^{\lambda} -1 + \beta} ds\\
&+ \int_{\sigma}^{\lambda} \frac{e^s}{e^{\lambda}-1+\beta}ds + (1)\frac{\beta}{e^{\lambda}-1+\beta}\displaybreak[0]\\
&= \int_0^{\lambda}\frac{e^s}{e^{\lambda}-1+\beta}ds +   \int_0^{\sigma} \Bigg[ \frac{1-\sigma + s}{\sigma} (1-\beta)\Bigg]\frac{e^s}{e^{\lambda} -1 + \beta} ds\\ &+\frac{\beta}{e^{\lambda}-1+\beta} \displaybreak[1]\\
&= \frac{e^{\lambda}-1}{e^{\lambda}-1+\beta} + \frac{1-\beta}{e^{\lambda}-1+\beta} + \frac{\beta}{e^{\lambda}-1+\beta}\displaybreak[2]\\
&= \frac{e^{\lambda}}{e^{\lambda}-1+\beta}.
\end{align*}

(ii) $\hat{\sigma} >1, \lambda \leq \sigma < 1$.  Note this is also a worst case failed prediction scenario.
\begin{align*}
&\int_s h(s,\sigma) f_1^*(s) ds\\ &= \int_0^{\lambda} \Bigg[ 1 + \frac{1-\sigma + s}{\sigma} (1-\beta)\Bigg]\frac{e^s}{e^{\lambda} -1 + \beta} ds + (1)\frac{\beta}{e^{\lambda}-1+\beta}\displaybreak[0]\\
&\leq \int_0^{\lambda}\frac{e^s}{e^{\lambda}-1+\beta}ds +   \int_0^{\sigma} \Bigg[ \frac{1-\sigma + s}{\sigma} (1-\beta)\Bigg]\frac{e^s}{e^{\lambda} -1 + \beta} ds +\frac{\beta}{e^{\lambda}-1+\beta} \displaybreak[1]\\
&= \frac{e^{\lambda}-1}{e^{\lambda}-1+\beta} + \frac{1-\beta}{e^{\lambda}-1+\beta} + \frac{\beta}{e^{\lambda}-1+\beta}\displaybreak[2]\\
&= \frac{e^{\lambda}}{e^{\lambda}-1+\beta}.
\end{align*}

(iii) $\hat{\sigma} >1, \lambda  < 1< \sigma$.  Note this is a best case correct prediction scenario.
\begin{align*}
&\int_s h(s,\sigma) f_1^*(s) ds \\
&= \int_0^{\lambda} \Bigg[  1 + \frac{s(1-\beta)}{(\sigma-1)\beta +1}\Bigg]\frac{e^s}{e^{\lambda} -1 + \beta} ds 
+ \Bigg[1 + \frac{(\sigma-1)(1-\beta)}{(\sigma-1)\beta +1}\Bigg]\frac{\beta}{e^{\lambda}-1+\beta}\\
&= \frac{1}{e^{\lambda} -1 + \beta} \Bigg\{ \int_0^{\lambda} e^s ds + \beta +\frac{1-\beta}{(\sigma-1)\beta +1} \Bigg[\int_0^{\lambda} se^s ds 
+ (\sigma -1) \beta \Bigg]\Bigg\} \nonumber\\
&= \frac{1}{e^{\lambda} -1 + \beta}  \Bigg\{ e^{\lambda} -1 + \beta + \frac{1-\beta}{(\sigma-1)\beta +1} \Bigg[e^{\lambda} (\lambda -1 ) + 1
+ (\sigma -1) \beta \Bigg]\Bigg\} \nonumber\displaybreak[0]\\
&= 1 + \frac{1}{e^{\lambda} -1 + \beta} \Bigg[ \frac{(1-\beta)e^{\lambda}(\lambda -1)}{(\sigma-1)\beta + 1} + (1- \beta) \Bigg] \nonumber\displaybreak[1]\\
&= \frac{e^{\lambda}}{e^{\lambda}-1+\beta} +  \frac{e^{\lambda}}{e^{\lambda}-1+\beta} \Bigg[\frac{(\lambda -1)(1-\beta)}{(\sigma-1)\beta + 1} \Bigg]\nonumber\displaybreak[2]\\
&= \frac{e^{\lambda}}{e^{\lambda}-1+\beta} -  \frac{e^{\lambda}}{e^{\lambda}-1+\beta} \Bigg[\frac{(1-\lambda)(1-\beta)}{(\sigma-1)\beta + 1} \Bigg]\nonumber\displaybreak[3]\\
&\leq \frac{e^{\lambda}}{e^{\lambda}-1+\beta}.
\end{align*}

(iv) $\hat{\sigma} \leq 1,  1 \leq 1/{\lambda} <  \sigma$.  Note this is a worst case failed prediction scenario.  
\begin{align*}
&\int_s h(s,\sigma) f_2^*(s) ds\\
&= \int_0^{1/
	{\lambda}} \Bigg[  1 + \frac{s(1-\beta)}{(\sigma-1)\beta +1}\Bigg]\frac{e^s}{e^{1/
		{\lambda}} -1 + \beta} ds 
+ \Bigg[1 + \frac{(\sigma-1)(1-\beta)}{(\sigma-1)\beta +1}\Bigg]\frac{\beta}{e^{1/
		{\lambda}}-1+\beta} \nonumber\displaybreak[0]\\
&= \frac{1}{e^{1/{\lambda}} -1 + \beta} \Bigg\{ \int_0^{1/
	{\lambda}} e^s ds + \beta +\frac{1-\beta}{(\sigma-1)\beta +1} \Bigg[\int_0^{1/
	{\lambda}} se^s ds  + (\sigma -1) \beta \Bigg]\Bigg\} \nonumber\displaybreak[1]\\
&= \frac{1}{e^{1/{\lambda}} -1 + \beta}  \Bigg\{ e^{1/{\lambda}} -1 + \beta 
+ \frac{1-\beta}{(\sigma-1)\beta +1} \Bigg[e^{1/
	{\lambda}} (1/{\lambda} -1 ) + 1  + (\sigma -1) \beta \Bigg]\Bigg\} \nonumber\displaybreak[2]\\
&= 1 + \frac{1}{e^{1/{\lambda}} -1 + \beta} \Bigg[ \frac{(1-\beta)e^{1/{\lambda}}(1/{\lambda} -1)}{(\sigma-1)\beta + 1} + (1- \beta) \Bigg] \nonumber\displaybreak[3]\\
&= \frac{e^{1/{\lambda}}}{e^{1/{\lambda}}-1+\beta} +  \frac{e^{1/{\lambda}}}{e^{1/{\lambda}}-1+\beta} \Bigg[\frac{(1/{\lambda} -1)(1-\beta)}{(\sigma-1)\beta + 1} \Bigg]\nonumber\\
&= \frac{e^{1/{\lambda}}}{e^{1/{\lambda}}-1+\beta} \Bigg[1+\frac{(1/{\lambda} -1)(1-\beta)}{(\sigma-1)\beta + 1} \Bigg]\nonumber\\
&\stackrel{(c)}{\leq} \frac{e^{1/{\lambda}}}{e^{1/{\lambda}}-1+\beta} \Bigg[1+(1/{\lambda} -1)(1-\beta) \Bigg] \\\displaybreak
&\stackrel{(d)}{\leq} \frac{e^{1/{\lambda}}}{e^{1/{\lambda}}-1+\beta} \Bigg[1+(1/{\lambda} -1) \Bigg]\\
&= \frac{1}{\lambda}\frac{e^{1/{\lambda}}}{e^{1/{\lambda}}-1+\beta},\nonumber
\end{align*}
where (c) holds since $(\sigma -1)\beta +1 \geq 1$, and (d) is true since $0 \leq 1-\beta \leq 1$.  However, note the following upper bound also holds
\begin{align*}
&1+\frac{(1/{\lambda} -1)(1-\beta)}{(\sigma-1)\beta + 1} \leq 1+\frac{(\sigma -1)(1-\beta)}{(\sigma-1)\beta + 1}\\ 
&\leq 1+\frac{(\sigma -1)(1-\beta)}{(\sigma-1)\beta } 
\leq 1+\frac{(1-\beta)}{\beta } 
\leq \frac{1}{\beta}.
\end{align*}
Then the competitive ratio in this case is
$$\int_s h(s,\sigma) f_2^*(s) ds \leq \min \Big\{1/\beta, 1/\lambda\Big\} \cdot \frac{e^{1/{\lambda}}}{e^{1/{\lambda}}-1+\beta}.$$

(iv) $\hat{\sigma} \leq 1,  1 \leq \sigma <  1/{\lambda}$.  Note this is a worst case failed prediction scenario.  
\begin{align*}
&\int_s h(s,\sigma) f_2^*(s) ds \\
&= \int_0^{\sigma} \Bigg[  1 + \frac{s(1-\beta)}{(\sigma-1)\beta +1}\Bigg]\frac{e^s}{e^{1/
		{\lambda}} -1 + \beta} ds 
+ \Bigg[1 + \frac{(\sigma-1)(1-\beta)}{(\sigma-1)\beta +1}\Bigg]\Bigg[\int_{\sigma}^{1/{\lambda}}\frac{e^s}{e^{1/
		{\lambda}} -1 + \beta}ds 
+ \frac{\beta}{e^{1/
		{\lambda}}-1+\beta}\Bigg]\nonumber\displaybreak[0]\\
&= \frac{1}{e^{1/{\lambda}} -1 + \beta} \Bigg\{ \int_0^{1/
	{\lambda}} e^s ds + \beta +\frac{1-\beta}{(\sigma-1)\beta +1} \Bigg[\int_0^{\sigma} se^s ds 
+ (\sigma -1)\int_{\sigma}^{1/{\lambda}}e^s ds + (\sigma -1) \beta \Bigg]\Bigg\} \nonumber\\
&=\frac{1}{e^{1/{\lambda}} -1 + \beta}  \Bigg\{ e^{1/{\lambda}} -1 + \beta + \frac{1-\beta}{(\sigma-1)\beta +1} \Bigg[e^{\sigma} (\sigma -1 )
+ 1  + (\frac{1}{\lambda} -1)e^{1/{\lambda}} -(\sigma-1) e^{\sigma} + (\sigma-1)\beta )\Bigg]\Bigg\} \nonumber\displaybreak[1]\\
&= 1 + \frac{1}{e^{1/{\lambda}} -1 + \beta}  \frac{1-\beta}{(\sigma-1)\beta +1} \Bigg[(\sigma -1 )\beta + 1 \displaybreak[2]
+ (1/{\lambda} -1)e^{1/{\lambda}}\Bigg]\nonumber\displaybreak[3]\\
&= 1 + \frac{1}{e^{1/{\lambda}} -1 + \beta} \Bigg[ (1-\beta) + \frac{(1/{\lambda}-1)(1-\beta)e^{1/{\lambda}}}{(\sigma-1)\beta+1}\Bigg] \nonumber\\
&= \frac{e^{1/{\lambda}}}{e^{1/{\lambda}}-1+\beta} +  \frac{e^{1/{\lambda}}}{e^{1/{\lambda}}-1+\beta} \Bigg[\frac{(1/{\lambda} -1)(1-\beta)}{(\sigma-1)\beta + 1} \Bigg]\nonumber\\
&= \frac{e^{1/{\lambda}}}{e^{1/{\lambda}}-1+\beta} \Bigg[1+\frac{(1/{\lambda} -1)(1-\beta)}{(\sigma-1)\beta + 1} \Bigg]\nonumber\\
&= \frac{e^{1/{\lambda}}}{e^{1/{\lambda}}-1+\beta} \frac{(\sigma-1/{\lambda})\beta+\frac{1}{\lambda}}{(\sigma-1)\beta + 1} \nonumber\\
&\stackrel{(d)}{\leq}\frac{e^{1/{\lambda}}}{e^{1/{\lambda}}-1+\beta}\frac{1/{\lambda}}{1}\nonumber\\
&=\frac{1}{\lambda}\frac{e^{1/{\lambda}}}{e^{1/{\lambda}}-1+\beta},
\end{align*}
where (d) is true since $1\leq\sigma\leq 1/{\lambda}$, $\sigma-1/{\lambda}<0$, and $\sigma-1\geq0.$
Then the competitive ratio in this case is
$$\int_s h(s,\sigma) f_2^*(s) ds \leq \min \Big\{1/{\beta }, 1/{\lambda}\Big\} \cdot \frac{e^{1/{\lambda}}}{e^{1/{\lambda}}-1+\beta}.$$

(vi) $\hat{\sigma} \leq 1,  \sigma \leq 1 <  1/{\lambda}$.  Note this is a best case correct prediction scenario.  
\begin{align*}
&\int_s h(s,\sigma) f_2^*(s) ds \\
&= \int_0^{\sigma} \Bigg[ 1 + \frac{1-\sigma + s}{\sigma} (1-\beta)\Bigg]\frac{e^s}{e^{1/{\lambda}} -1 + \beta} ds 
+ \int_{\sigma}^{1/{\lambda}} \frac{e^s}{e^{1/{\lambda}}-1+\beta}ds + (1)\frac{\beta}{e^{1/{\lambda}}-1+\beta}\\
&= \int_0^{1/{\lambda}}\frac{e^s}{e^{1/{\lambda}}-1+\beta}ds +   \int_0^{\sigma} \Bigg[ \frac{1-\sigma + s}{\sigma} (1-\beta)\Bigg]
\cdot\frac{e^s}{e^{1/{\lambda}} -1 + \beta} ds +\frac{\beta}{e^{1/{\lambda}}-1+\beta} \\
&= \frac{e^{1/{\lambda}}-1}{e^{1/{\lambda}}-1+\beta} + \frac{1-\beta}{e^{1/{\lambda}}-1+\beta} + \frac{\beta}{e^{1/{\lambda}}-1+\beta}\\
&= \frac{e^{1/{\lambda}}}{e^{1/{\lambda}}-1+\beta}.
\end{align*}

Next, we consider the consistency. 
For consistency guarantees, we compute the competitive ratio assuming the predictions are correct.  There are two cases to consider here

(i) $\hat{\sigma} = \sigma >1$.  With a selected parameter $s$ from the distribution $ f_1^*(s)$, the algorithm uses the local generator for the first $T^s$ time slots before switching to the grid.  Then the cost of the algorithm is  $\texttt{ALG}= \sum_{t=1}^{T^s}p_gd(t) + \sum_{t=T^s}^{T}p(t)d(t) + p_m$.  Since $\sigma > 1$, the optimal offline solution uses the grid for the whole duration with cost  $\texttt{OPT}= \sum_{t=1}^{T}p(t)d(t) + p_m$.  Then we have the following:
\begin{align*}
\texttt{ALG} &=\sum_{t=1}^{T^s}p_gd(t) + \sum_{t=T^s+1}^{T}p(t)d(t) + p_m\\
&=\sum_{t=1}^{T^s}(p_g - p(t))d(t) + \sum_{t=1}^{T}p(t)d(t) + p_m\\
&\leq s\cdot p_m + \sum_{t=1}^{T}p(t)d(t) + p_m\\
&\leq(1+s) \cdot p_m + \sum_{t=1}^{T}p(t)d(t) \\
&\leq(1+s) (p_m + \sum_{t=1}^{T}p(t)d(t)) 
\leq(1+s)\texttt{OPT}. \\
\end{align*}

To compute the expected expected cost of the randomized algorithm, we need to know a special case of the cost of \texttt{ALG} when $s=\infty$.  With $s=\infty$, the algorithm never switches to grid electricity
\begin{align*}
\texttt{ALG}_{\{s = \infty\}} &=\sum_{t=1}^{T}p_gd(t) 
=\sum_{t=1}^{T}\frac{p_g}{p(t)}p(t)d(t) \\
&\leq\frac{p_g}{p^{min}}\sum_{t=1}^{T}p(t)d(t) 
=\frac{1}{\beta}\sum_{t=1}^{T}p(t)d(t)\displaybreak[0]\\
&\leq\frac{1}{\beta}\sum_{t=1}^{T}p(t)d(t) + \frac{1}{\beta}p_m
=\frac{1}{\beta}\texttt{OPT}.
\end{align*}

Then the expected cost of the randomized algorithm is

\begin{align*}
&\mathbb{E}[\texttt{ALG}] \\&= \int_s \texttt{ALG}\cdot  f_1^*(s) ds\\
&\leq \int_0^{\lambda} (1+s) (\texttt{OPT}) \frac{e^s}{e^{\lambda}-1+\beta} ds + \frac{1}{\beta}(\texttt{OPT}) \frac{\beta}{e^{\lambda}-1+\beta}\displaybreak[0]\\
&\leq \frac{\texttt{OPT}}{e^{\lambda}-1+\beta}\Bigg[1+\int_0^{\lambda} e^s + se^s ds\Bigg]
= \frac{\texttt{OPT}}{e^{\lambda}-1+\beta}(1+\lambda e^{\lambda}).
\end{align*}
If $\lambda=0$, we have $(1/\beta)$-consist.

(ii) $\hat{\sigma} = \sigma \leq 1$.  With hyperparameter $\lambda$, the algorithm uses the local generator for the first $T^{1/{\lambda}}$ time slots before switching to the grid, where $T^{1/{\lambda}} \leq T$.  Then the cost of the algorithm is  $\texttt{ALG}= \sum_{t=1}^{T^{1/{\lambda}}}p_gd(t) + \sum_{t=T^{1/{\lambda}}+1}^{T}p(t)d(t) + p_m$.  Since $\sigma \leq 1$, the optimal offline solution uses the grid for the whole duration with cost  $\texttt{OPT}= \sum_{t=1}^{T}p_gd(t) $.  Then we have the following:
\begin{align*}
\texttt{ALG} &= \sum_{t=1}^{T^{1/{\lambda}}}p_gd(t) + \sum_{t=T^{1/{\lambda}}+1}^{T}p(t)d(t) + p_m\displaybreak[0]\\
&\leq \sum_{t=1}^{T^{1/{\lambda}}}p_gd(t) + \sum_{t=T^{1/{\lambda}}+1}^{T}p_gd(t) + p_m
= \texttt{OPT} + p_m\displaybreak[1]\\
&\stackrel{(e)}{\leq} \texttt{OPT} + \lambda \Bigg(\sum_{t=1}^{T^{1/{\lambda}}+1}(p_g - p(t))d(t)\Bigg)\\
&\leq \texttt{OPT} + \lambda \Bigg(\sum_{t=1}^{T}p_gd(t)\Bigg)
= (1+\lambda)\texttt{OPT}, \\
\end{align*}
where (e) is true from Algorithm \red.

\end{document}